\documentclass[prc,aps,amsfonts,amsmath,superscriptaddress,floatfix,twocolumn]{revtex4}
\usepackage{graphicx,float} 
\usepackage{epsfig} 
\usepackage{rotating}
\usepackage{mwe}    
\usepackage{subfig}

\begin{document}

\title{Localisation and dimensionless studies in nuclei and many-body systems}

\author{J.-P. Ebran}
\affiliation{CEA,DAM,DIF, F-91297 Arpajon, France}
\author{E. Khan}
\affiliation{Institut de Physique Nucl\'eaire, Universit\'e Paris-Sud, IN2P3-CNRS, 
Universit\'e Paris-Saclay, F-91406 Orsay Cedex, France}

\begin{abstract} 

Dimensionless ratios characterizing many-body systems are a powerful tool to reveal the main universal quantities involved.
The recently-introduced localisation parameter allow to study the occurrence of crystal, clusterisation, and quantum liquid states in many-body systems such as nuclei.
Its concomitant use with other dimensionless quantities such as the quantality, allows to pinpoint fundamental lengthscales and dimensionless ratios at work in nuclei. Within the present approach, the impact of delocalisation on nuclear saturation is discussed. The transition from a homogeneous to a clusterised state in low density systems is studied. The spin-orbit effect in nuclei and fermionic systems is also reviewed in the light of this analysis. A generalised localisation parameter is derived, showing that the delocalisation properties are not always driven by the quantality. The concepts of quantum fluidity and mobility are finally introduced.

\end{abstract}
 


\date{\today}

\maketitle

\section{Introduction}

Localisation studies in many-body systems provide a useful insight on their general behavior \cite{joh15,nan15}. In nuclear systems, such studies addressed clusterisation, quantum liquid (QL) or fission issues \cite{zin07,hor10, ebr12,zin13,ebr13,ebr14a,ebr14,naz16},  showing a growing interest for this powerful approach, and yielding a large variety of results.
Identifying the fundamental parameters driving localisation in fermionic finite systems, especially in nuclei, is therefore of broad interest. Dimensionless ratios
are efficient tools to perform such an analysis. Several studies used dimensionless ratios in order to investigate many-body
systems \cite{bon03,zin13}, providing useful results on the behavior of these systems. For instance the quantality \cite{mot96} is a dimensionless ratio used in a few works \cite{zin13,ebr13} and it is timely to clarify the respective roles of quantality and localisation, especially in nuclei. Another dimensionless ratio is the localisation parameter introduced in our previous works and allowing to describe transition from the homogenous to cluster states in finite systems, in agreement with microscopic predictions \cite{ebr12,ebr13,ebr14a}.

Localisation is also related to fundamental properties of nuclei, such as saturation. It is well-known that the saturation density parameter (i.e. the nucleon-nucleon distance) is about 1.2 fm whereas phase shift analyses show that the hard-core of the nucleon-nucleon interaction, responsible for saturation, extents to about 0.5 fm \cite{bm69}. The nuclear saturation density (0.16 fm$^{-3}$) is consequently about one order of magnitude smaller than the packed density which would corresponds to nucleons in hard-core contact from each other.
Some delocalisation effects could explain this feature. Delocalisation drives the QL versus crystaline properties of a system. While discussing the origin of saturation in
nuclei, A. Bohr and B. Mottelson state: "The nature of the transition from
independent-particle motion to the crystalline state and the
associated value of the characteristic parameter present significant
unsolved problems" \cite{bm69}. The essential role of delocalisation on nuclear saturation
have surprisingly not been tackled yet, although it is obvious that 
a nuclear cluster state such as the Hoyle one may not exhibit saturation density in 
its very core \cite{hor10,rop09,sch13,ebr14}. Localisation studies shall play a pivotal role, not only in 
describing the QL to cluster transition, but also into establishing the magnitude of the saturation density
in nuclear states. 

It may therefore be relevant to take the full advantage of approaches using dimensionless ratios (including  the localisation parameter) in the most fundamental and systematic way, in order to deeply understand localisation and other underlying effects, especially in nuclear matter and nuclei. The fine structure constant \cite{som}, or its generalisation to
non-electromagnetic interactions, is of course one of the most
useful dimensionless quantity. However it only characterizes the
interaction at work in the system, and additional dimensionless
quantities are required in order to fully describe an interacting
system of massive constituents. Another relevant dimensionless quantity is the above-mentioned quantality
\cite{boe48}. It allows to characterize either the crystal or the
QL behavior of a system. Although derived several
decades ago, it has been rather recently introduced in nuclear physics
\cite{mot96}, and only a few works have explored its impact on nuclei
\cite{zin07,ebr12,zin13,ebr13,ebr14a,ebr14}. The recently introduced localisation parameter in nuclei \cite{ebr13} may also be a relevant missing piece for localisation and dimensionless analyses. It allows to analyse the transition from the homogeneous to clusterised density in nuclei and finite systems. Another dimensionless quantity has recently been established, related to the spin-orbit effect in many-body fermionic systems \cite{ebr15}. The concomitant use of
these dimensionless quantities (the coupling constant, the quantality, the localisation parameter and the spin-orbit related one) and the investigation of their
relation has therefore never been undertaken so far. Such a universal approach shall bring a useful light on localisation in many-body systems with finite-size effects and on nuclear saturation. It could
potentially lead to fundamental predictions for finite systems by considering localisation in its most general way. 

The full knowledge which
can be extracted from the quantality is therefore in order, as well as going one step further by considering the localisation parameter, which is designed for finite systems. Section II introduces fundamental dimensionless quantities such as the localisation parameter, the quantality and the inertia parameter. The corresponding
discussion on the spatial dispersion of constituents and relevant lenthscales is undertaken. In section III the impact of nucleonic delocalisation on the nuclear saturation density is studied. The occurrence of a cluster phase in low density nuclear systems is also discussed using dimensionless quantities. Section IV provides several other applications of such a use of dimensionless studies, allowing for a renewed light on the spin-orbit effect and a generalisation of the localisation parameter. Finally a systematic derivation of dimensionless quantities in many-body systems is undertaken, introducing the quantum fluidity and mobility. Appendix A discusses the relation between the reduced Compton wavelength and the quantities introduced in the present work. Appendix B provides hints on how the present approach could be applied to the case of graphene.

\section{Localisation, quantality and lengthscales}

\subsection{Localisation and clustering}

In finite systems where surface effects are not a perturbation, such
as in nuclei, it has been shown that a relevant quantity to describe the emergence of localisation effects
is the localisation parameter \cite{ebr13}. It allows for instance
to describe clusters states in nuclei, which can be considered as hybrid
states between crystals and quantum liquids. 

The localisation parameter $\alpha_{loc}$ is defined as the ratio of the typical size
of the constituent wave function (in the finite system) to the interconstituent distance. 
This parameter is very useful to study the occurrence of clusterisation in finite systems such as nuclei \cite{ebr12,ebr13,ebr14a}, and the transition from homogeneous to more spatially localised densities. 
In an Harmonic Oscillator (HO) approximation of the confining potential, the analytic expression for the localisation
parameter $\alpha_{loc}$ is \cite{ebr13}:

\begin{equation}
\alpha_{loc}\equiv\frac{b}{r_0}=\frac{\sqrt{\hbar}A^{1/6}}{(2m_NV_0r_0^2)^{1/4}}
\label{eq:alpha}
\end{equation}

where V$_0$ is the depth (V$_0$ $>$ 0) of the confining mean-field 
potential, A the number of constituents of the system and b the typical spread of the wavefunction of a constituent in the finite
system. The above equation holds for a system where
R$\simeq$r$_0$A$^{1/3}$ with r$_0$ the typical interconstituent lengthscale. 
The localisation parameter is $\alpha_{loc}\simeq$1 for the hybrid
(cluster) phase between the QL ($\alpha_{loc}>$1) and the crystal
($\alpha_{loc}<$1) one \cite{ebr12}. For instance, Eq.
(\ref{eq:alpha}) yields $\alpha_{loc}\simeq$1.5 for an A $\sim$ 100
nucleus, corresponding to an homogeneous density (QL regime). The analytical expression (\ref{eq:alpha}) of the localisation parameter
 is a good approximation of the localisation parameter directly obtained 
from a fully microscopic calculation (see Ref. \cite{ebr13} for more details).

Localisation effects can be investigated from a microscopic point of view within, e.g., the nuclear energy density functional (EDF) approach. The basic rationale of the EDF approach is to split nucleonic correlations into different categories:
\begin{enumerate}
 \item[i)] the bulk of correlations, varying smoothly with the number of nucleons, is implicitly resummed in a phenomenological energy functional. This allows to deal with the ultraviolet nonperturbative nature of the nuclear manybody problem induced by the strong internucleon repulsion at short distance;
 \newline
 
  \item[ii)] strong nondynamical correlations involved in open-shell nuclei (angular and pairing correlations) are explicitly treated at a 'mean-field' level, hence ensuring a favourable numerical cost, by allowing the product state that approximates the total many-body state to spontaneously break symmetries of the system. A large part of the nonperturbative physics of infrared nature is then efficiently incorporated in a single product state; 
 \newline
 
  \item[iii)] additional nondynamical and dynamical nucleonic correlations are accounted for in a 'beyond mean-field' step where the A-body state representing the many-body system is expressed in terms of a continuous mixing of symmetry-breaking mean-field states.   
\end{enumerate}
$\phantom{nl}$
\newline

\begin{figure}[]  
      {\includegraphics[width=0.27\textwidth]{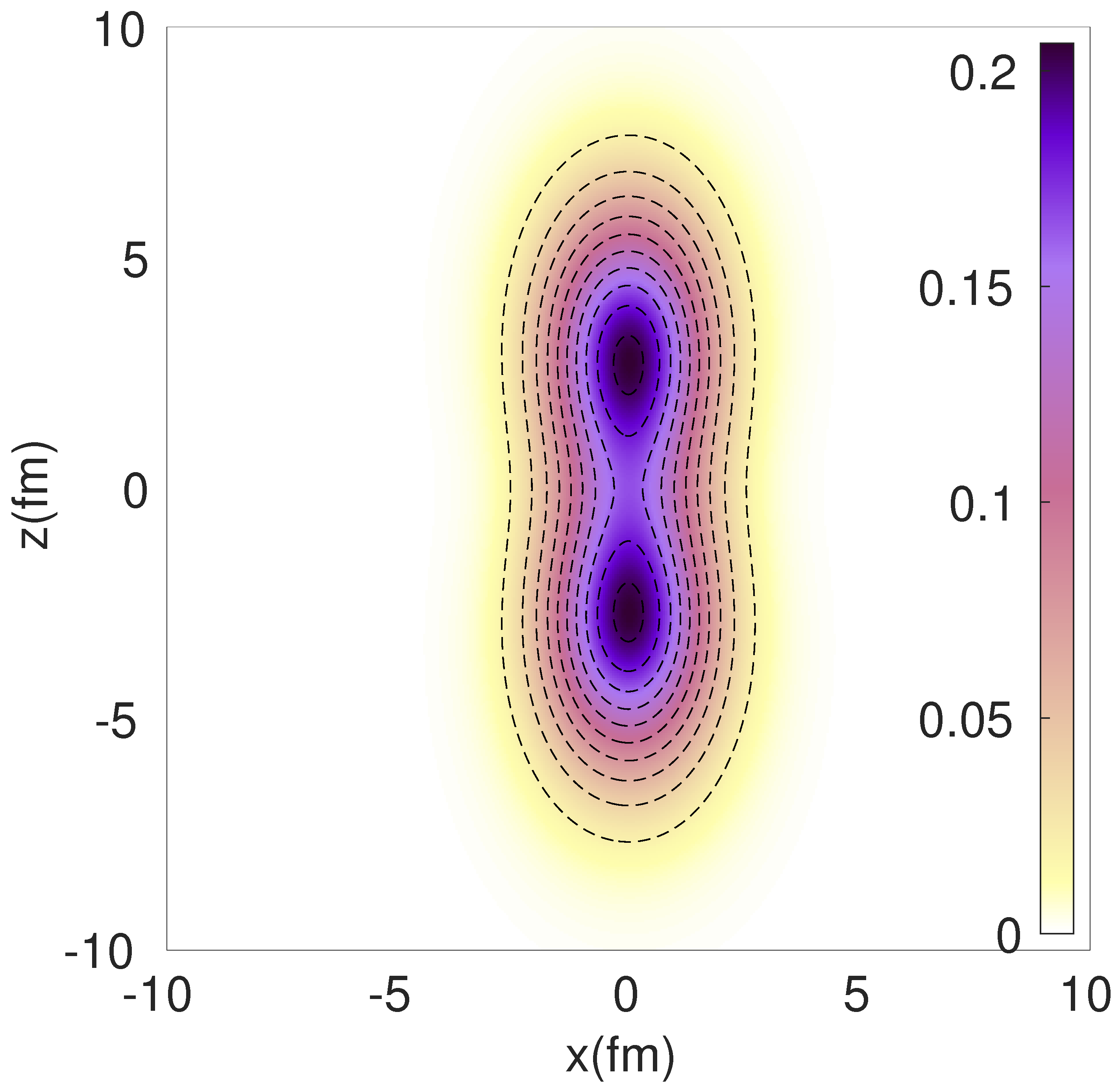}
      \includegraphics[width=0.27\textwidth]{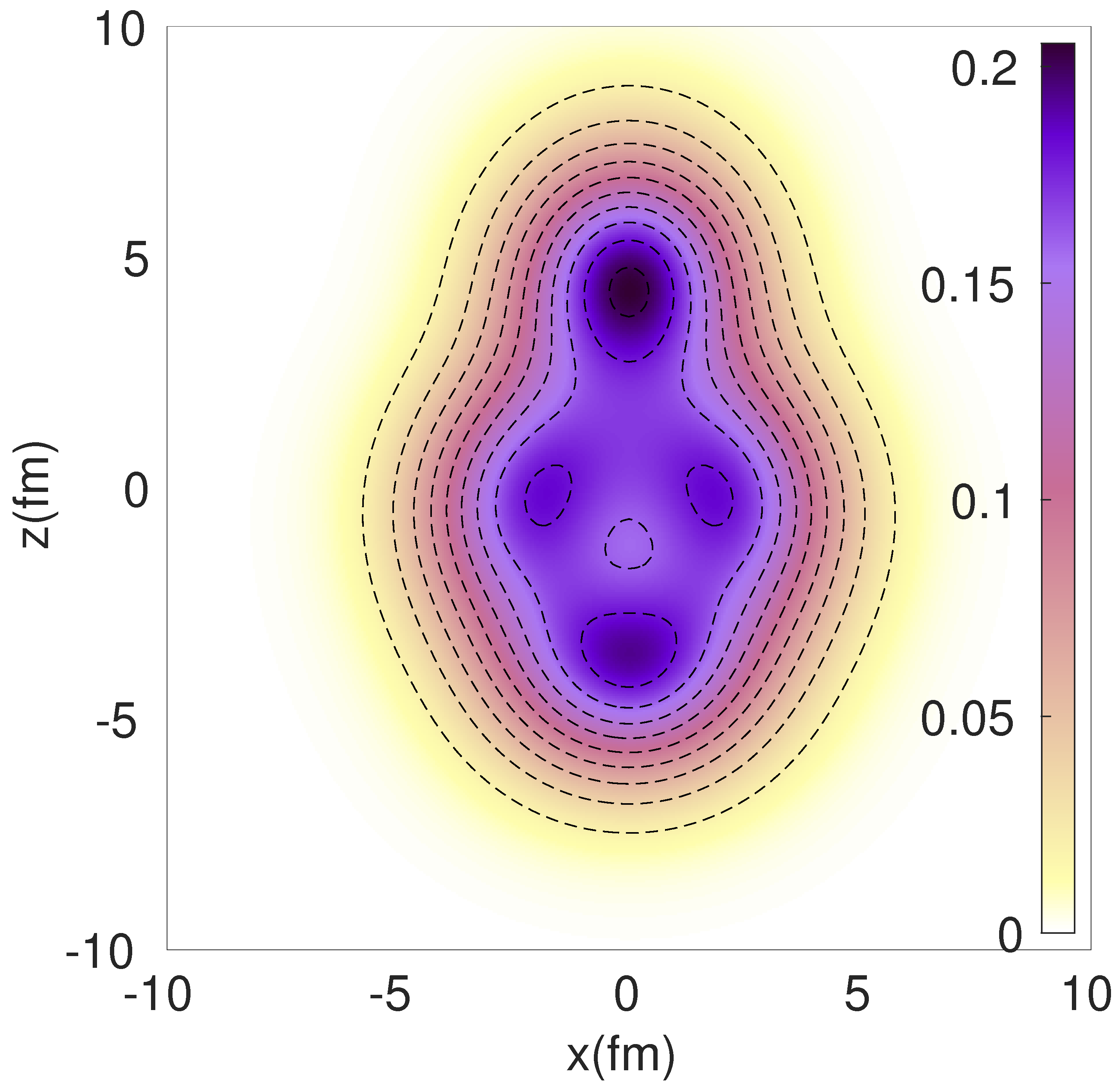}
      \includegraphics[width=0.27\textwidth]{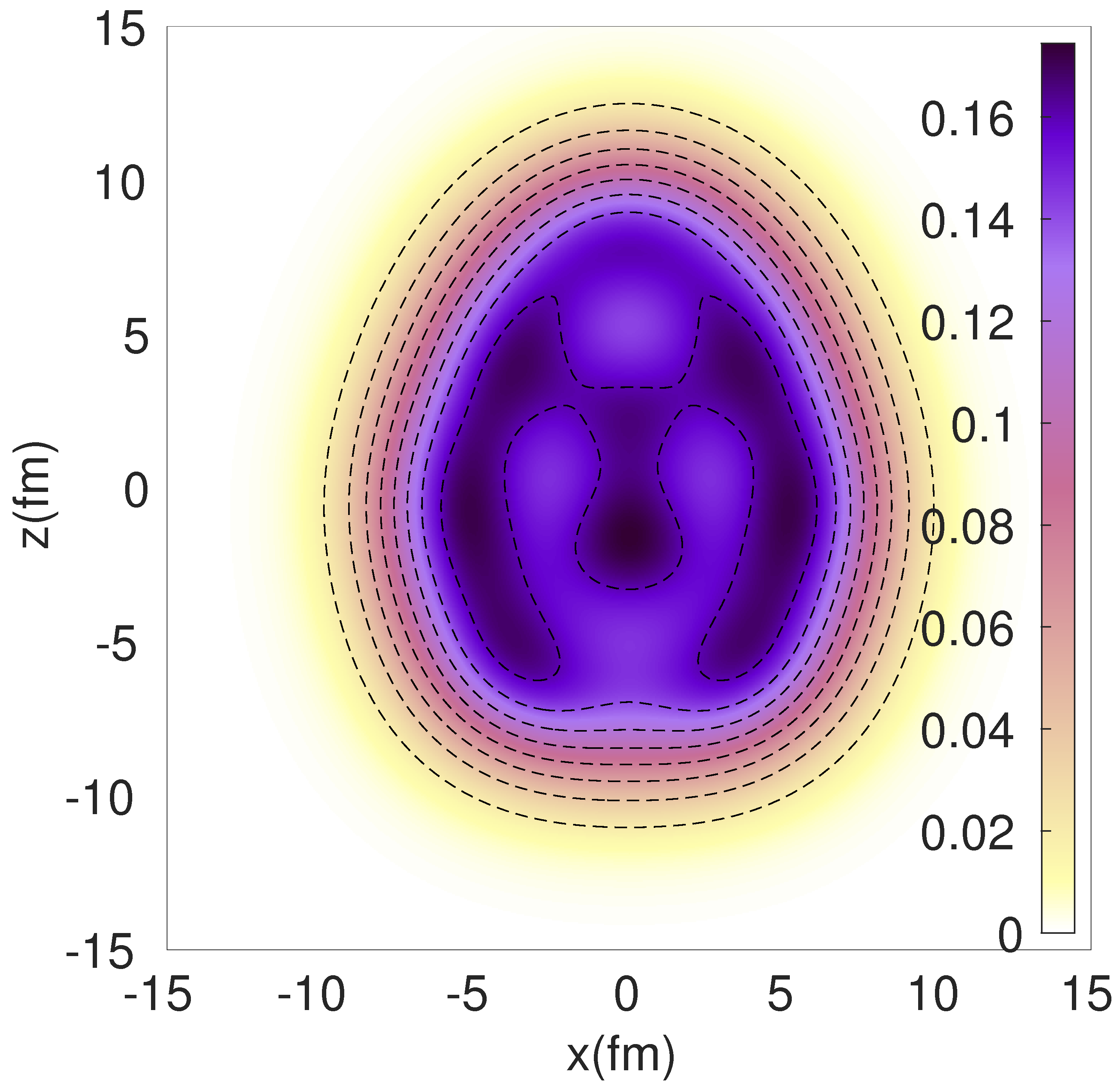}
      \includegraphics[width=0.27\textwidth]{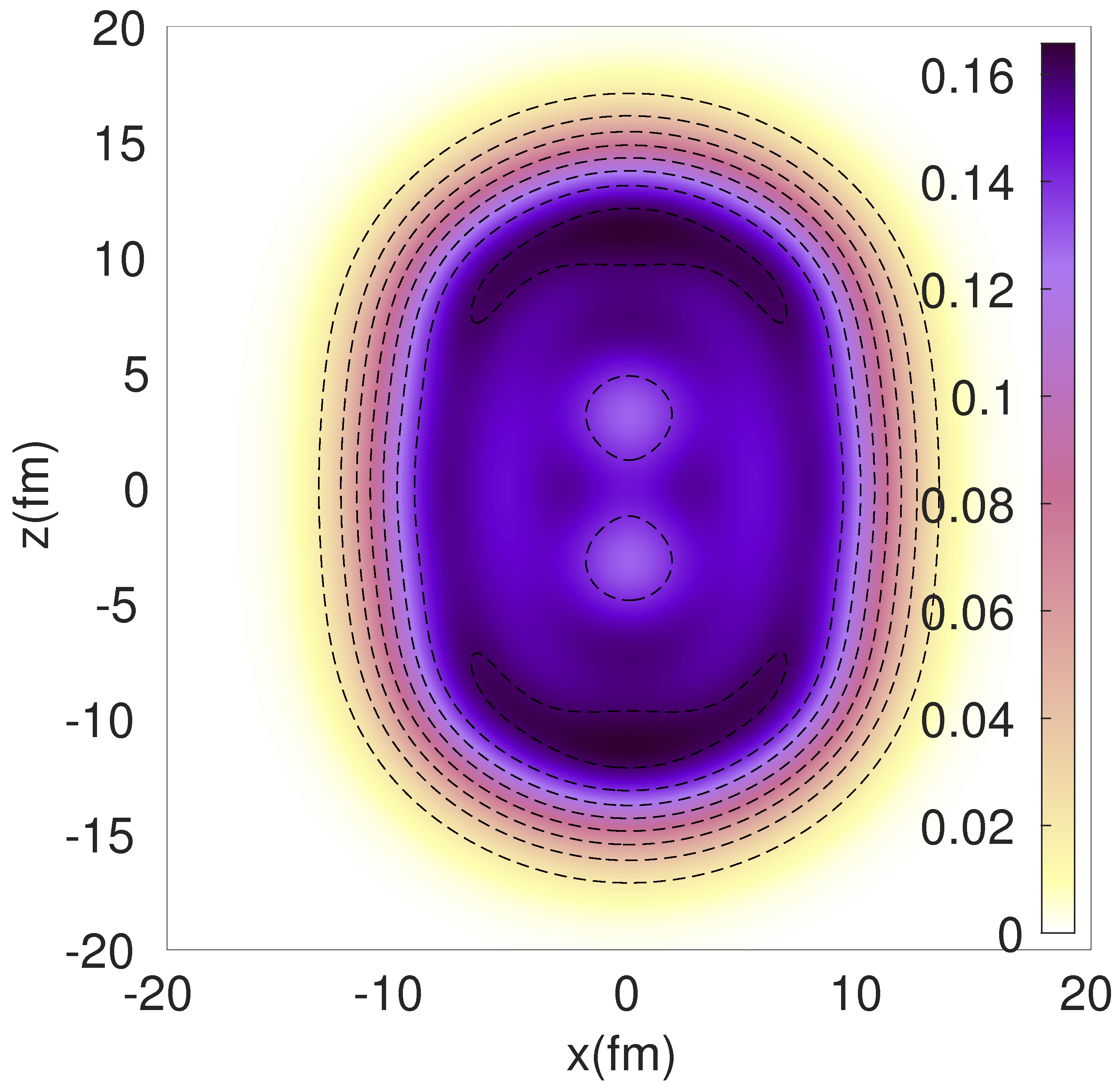}
      \includegraphics[width=0.27\textwidth]{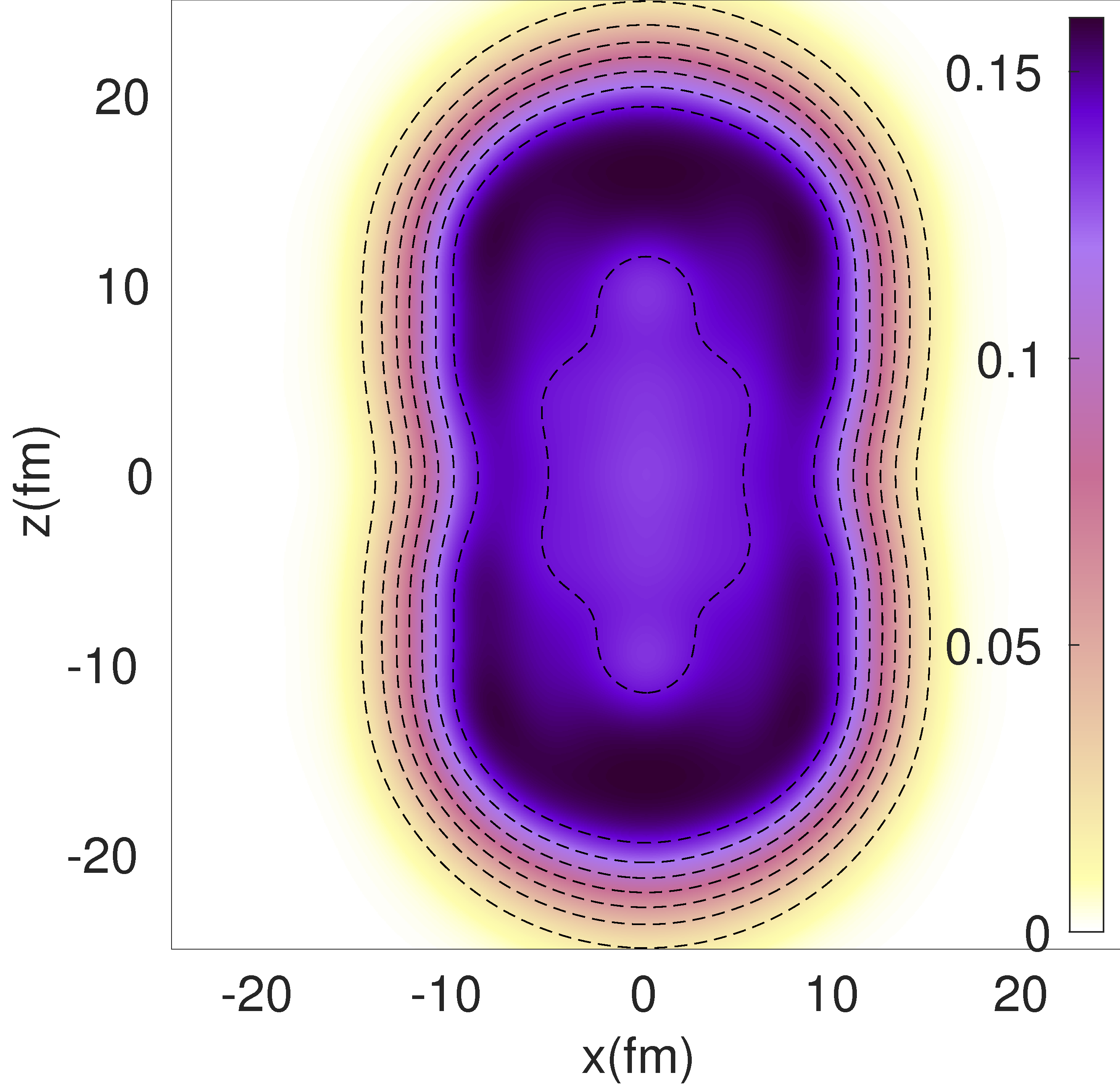}
    }
    \caption{(Color online) Octupole-quadrupole RHB calculations of $^{8}$Be, $^{20}$Ne, $^{62}$Zn,$^{130}$Nd,$^{250}$Cf total densities (from top to bottom) using the DDME2 functionnal}    
    \label{fig:mic}
  \end{figure}

The EDF calculations performed in the present work are based on the covariant functional DD-ME2~\cite{lala}. The symmetry under the global gauge group U(1) as well as the rotational and reflection symmetries are broken, such that pairing and deformation physics are embedded in a single product state. The corresponding Relativistic Hartree Bogoliubov (RHB) mean-field equations are solved under constraints specifying the extent to which symmetries are broken (quadrupole deformation parameter; octupole deformation parameter) \cite{vre05,nik14}. The reflection symmetry is restored in a beyond mean-field step. Fig~\ref{fig:mic} displays the ground-state intrinsic total density of $^{8}$Be, $^{20}$Ne, $^{62}$Zn, $^{130}$Nd and $^{250}$Cf in the (Oxz) plane. For the shape (elongation, reflection-asymmetry) that minimizes the energy of each nuclei, the EDF calculation gives the optimal repartition of the nucleons within the corresponding nuclear volume. In the lightest systems, arranging nucleons into alpha-type clusters is energetically favoured, whereas a clear dilution of the localisation pattern is observed ranging from light to heavy nuclei, in agreement with the localisation parameter predictions: the localisation
parameter $\alpha_{loc}$ increases with A, while V$_0$ does not change significantly from light to heavy nuclei because of the saturation effect. For a detailed discussion, see Refs. \cite{ebr13, ebr14a}.

The localisation parameter of Eq. (\ref{eq:alpha}) has been shown to be related to the quantality \cite{ebr14}. Before investigating this path in more details in section III, fundamental dimensionless ratios at work have to be introduced.

\subsection{Minimal universal information in a many-body system and dimensionless ratios}

From a general picture, the information of an interacting many-body system can be reduced to 3
basic quantities: i) the typical magnitude -V'$_0$ (V'$_0$ $>$0) of the two body
interaction, ii) its typical equilibrium lengthscale $a$ in the system and iii) the mass m$_N$ of each
interacting particle. It should be noted that the lengthscale $a$ corresponds to the range at equilibrium of the interaction. 
It should not be confused with the interconsituent distance r$_0$ (the density parameter) which can be significantly larger. 
In nuclei, phase shift analyses give $a$ $\simeq$ 0.9 fm \cite{mot96} whereas r$_0$=1.2 fm. The origin and implications of this difference is strongtly connected with the magnitude of the saturation density and will be discussed in section III.A. In a second step, the $a$ $\sim$ r$_0$ approximation will then be used in the following of this work.

In order to study universal features of many body systems, a powerful
tool is to build dimensionless ratios from V'$_0$, $a$ and m$_N$.  Among all the possible dimensionless ratios,
three of them have a specific meaning : the dimensionless
coupling constant $\alpha$, the so-called inertia parameter $\eta$
and the quantality $\Lambda$. The knowledge of $\alpha$ and $\eta$
are sufficient to characterize the system, and therefore $\Lambda$ can
be deduced from these two quantities.

\subsubsection{The coupling constant}

A first dimensionless ratio that can be defined is

\begin{equation}
\alpha\equiv\frac{aV'_0}{\hbar c}
\label{cc}
\end{equation}

This quantity only depends on the interaction and not on the
constituent. It can be interpreted as the coupling constant of the
interaction: in the case of the electromagnetic interaction, Eq.
(\ref{cc}) gives

\begin{equation}
\alpha_{EM}\simeq\frac{e^2}{4\pi\epsilon_0\hbar c}=\frac{1}{137}
\label{cem}
\end{equation}

which is of course the fine structure constant.

In the case of the strong interaction in nuclei, (\ref{cc}) yields

\begin{equation}
\alpha_{S}\simeq\frac{100 MeV. 1fm}{200 MeV.fm}\simeq 0.5-1
\label{cstr}
\end{equation}

which is the typical magnitude of the strong interaction in these
systems. In many-body systems, the effective interaction can take
various coupling constant values, such as in graphene where
$\alpha$=2.5 \cite{shy07} (see Appendix B).

\subsubsection{The inertia parameter}

The inertia parameter $\eta$ is defined as

\begin{equation}
\eta\equiv\frac{m_Nc^2}{V'_0} 
\label{eta}
\end{equation}

If the typical kinetic energy T$_N$ of a constituent is approximated
to V'$_0$ in Eq. (\ref{eta}) then $\eta$ measures the relativistic effects at work in the
system: it is non-relativistic when $\eta$ goes to infinity,
relativistic when $\eta$ is close to 1, and ultra-relativistic when
$\eta$ goes to 0. 

It has been shown that in finite fermionic systems, the inertia parameter $\eta$ also drives the
relative magnitude of the spin-orbit (LS) effect (interestingly also of relativistic origin) not only in atoms but also
in other systems such as nuclei, hypernuclei and quarkonia
\cite{ebr15}. In atoms, $\eta$ is very large, leading to a large main shell
gap with respect to the LS one (fine structure). In nuclei, approximating V'$_0$ to typically the order of a hundred
MeV makes $\eta$ of the order of
few units and so does the HO gap with respect to the spin-orbit one.
The inertia parameter also allows to understand in a deeper way nuclear magicity
(see \cite{ebr15} for a more detailed discussion on the spin-orbit rule). A
giant spin-orbit state is also predicted for $\eta$=1/2, where the spin-orbit gap becomes
larger than the HO one. 

Formally, $\alpha$ and $\eta$ form the
complete set of dimensionless ratios which can built from the 3 main
quantities V'$_0$, $a$ and m$_N$ of a many-body system.  More advanced
dimensionless ratios can be built, and will be discussed in section IV.C: 
they can always be derived from $\alpha$ and $\eta$. Among them, the
quantality has a strong physical meaning.  

\subsubsection{The quantality}

The quantality $\Lambda$ is defined as the ratio of the zero point kinetic energy
T$_0$ to the magnitude of the interaction V'$_0$ \cite{mot96}:

\begin{equation}
\Lambda \equiv
\frac{T_0}{V'_0}\simeq\frac{\hbar^2}{m_Na^2V'_0}
\label{la}
\end{equation}

It is also built from V'$_0$, $a$ and m$_N$. T$_0$ should not be confused with the typical kinetic energy T$_N$ of a constituent:
T$_0$ is the kinetic energy contribution coming from quantum effects, and T$_N \ge $ T$_0$.

Mottelson used the quantality from de Boer \cite{boe48} to describe
when a system behaves like a quantum liquid rather than a
crystal. $\Lambda$ is large for QL states and small for crystal ones
\cite{mot96,zin13}. The typical value calculated with Eq. (\ref{la})
is $\Lambda \simeq$ 10$^{-2,-3}$ in the case of crystals like atoms
and molecules, and $\Lambda \simeq$ 0.1-1 (i.e. one order of magnitude larger) in the case of QL such as
$^4$He, nuclei or electrons in atoms. In the case of
nuclei, $\Lambda \simeq$ 0.5, using V'$_0$ $\simeq$ 100 MeV and r$_0$
$\simeq$ 1 fm \cite{mot96}.

It should be noted that the quantality carries similar information than $\cal{A}$, 
the action of the system normalised to $\hbar$:

\begin{equation}
{\cal A} \equiv \frac{a\sqrt{m_N.V'_0}}{\hbar}
\label{action}
\end{equation}

Eqs (\ref{la}) and (\ref{action}) yield

\begin{equation}
\Lambda=\frac{1}{{\cal A}^2}  
\label{lamact}
\end{equation}

It is well known that quantum effects in a system are large when its
action is close to $\hbar$ \cite{coh}. This corresponds to
$\cal{A}$$\gtrsim$ 1 and $\Lambda$ $\lesssim$ 1. This is indeed the
quantum liquid case, in agreement with the value of $\Lambda$
discussed above. When quantum effects are smaller, such as in the
crystal case, the action of the system is significantly larger than
$\hbar$: $\cal{A}$$\gg$ 1 and therefore $\Lambda$ $\ll$ 1: the present
use of the quantality and the action is relevant in order
to analyse the QL or crystal behavior of the system. Table
\ref{tab:quanta} summarizes
the various above discussed dimensionless quantities for several
many-body systems. It should be noted that i) this approach is valid for both fermions and bosons,
ii) the  quantum liquid nature of electron in atoms has been discussed in Ref. \cite{mot96},
and iii) results have to be considered within one order of magnitude for $\eta,\Lambda$ and
$\alpha$ due to the factorless definition of these quantities.

\begin{table*}[t]
\renewcommand{\arraystretch}{1.5}
\centering
\begin{tabular}{cccccccccccc}
\hline \hline
        Constituent  & m$_N$ & $a$ (nm) & V'$_0$ (eV) & T$_0$ (eV) 
	& $\alpha$ & $\eta$ & $\cal{A}$  &$\Lambda$ & State &   QF  & QMx3.10$^8$\\
\hline
$^{20}$Ne atom & 20 & 0.31 & 31 10$^{-4}$ & 2.2 10$^{-5}$  &4.9 10$^{-6}$ & 6.1 10$^{12}$&12.0& 0.007 & crystal  & 3.3 10$^{-8}$  & 1.6 10$^{-12}$   \\
H$_2$ molecule & 2 & 0.33  & 32 10$^{-4}$ & 1.9 10$^{-4}$ &5.3 10$^{-6}$& 5.9 10$^{11}$& 4.1 & 0.06 & crystal  & 3.2 10$^{-7}$ &  1.6 10$^{-10}$   \\
$^4$He atom & 4 & 0.29  & 8.6 10$^{-4}$ & 1.2 10$^{-4}$  & 1.2 10$^{-6}$ & 4.4 10$^{12}$& 2.7 & 0.14 & QL  &  1.9 10$^{-7}$  & 1.4 10$^{-11}$   \\
$^3$He atom & 3 & 0.29  & 8.6 10$^{-4}$ & 1.6 10$^{-4}$  &1.2 10$^{-6}$ & 3.2 10$^{12}$ & 2.3 & 0.19 & QL  &   2.6 10$^{-7}$  & 2.3 10$^{-11}$  \\
Nucleon & 1 & 9 10$^{-7}$ & 100 10$^{6}$ & 50 10$^{6}$ & 0.46 & 9.4 & 1.4 & 0.5 & QL   &   0.2    & 7.3 10$^{6}$ \\
e$^-$ in atoms & 5 10$^{-4}$ & 0.05 & 10 & 31 & 2.5 10$^{-3}$& 5 10$^{4}$&0.6 & 3.1 & QL  & 8.1 10$^{-3}$   &  47   \\
\hline \hline
\end{tabular}
\caption{Constituent mass, interaction lengthscale and magnitude, zero point kinetic energy, effective coupling constant, inertia parameter \cite{ebr15}, action and quantality \cite{mot96}
for various many-body systems. 
m$_N$ is given in units of a nucleon mass for convenience. The quantum fluidity (QF) and mobility (QM) are introduced in section IV.C.}
\label{tab:quanta}
\end{table*}

\subsubsection{The key relation}

Dimensionless quantities are known to be a powerful tool in physics \cite{uza} and
$\alpha$, $\eta$ and $\Lambda$ can be used to study phases of
many-body systems. The combination of these three ratios is sufficient to determine main behaviors of
many-body systems. $\alpha$ and $\eta$ form the
complete set of ratios to build any dimensionless quantity out of
V'$_0$, a and m$_N$. Eqs. (\ref{cc}), (\ref{la}) and
(\ref{eta}) yield a key relation among the three meaningful
dimensionless ratios:

\begin{equation}
\eta\Lambda\alpha^2=1
\label{key}
\end{equation}

which can also be written

\begin{equation}
\eta=\left(\frac{\cal A}{\alpha}\right)^2
\label{keyd2}
\end{equation}

This relation allows to make several physical interpretations. First, it is striking that setting the
interaction ($\alpha$) and the crystal vs. QL nature of the system
($\Lambda$) immediately yields the magnitude of the spin-orbit effect
($\eta$). Eq. (\ref{keyd2}) shows that the spin-orbit effect (in
finite systems) behaves as the coupling constant weighted by the action
of the system: a large magnitude of the spin-orbit (small $\eta$, see \cite{ebr15} and section IV.A) requires a large
coupling constant together with a small action. 

Nuclear states are known to behave as quantum liquid states and
therefore $\Lambda \lesssim$ 1, as displayed on Table I. Moreover, the low energy QCD
interaction yields $\alpha\sim$ 1. Eq. (\ref{keyd2}) then provides
$\eta\sim$ 1. This value of $\eta$ allows for strong spin-orbit effect
in finite systems \cite{ebr15}: m$_N$c$^2$ is of the order of the magnitude of the interaction V'$_0$ within one order of magnitude, as discussed
in section \ref{sec:ls}. The nucleus is therefore a very specific system where
$\eta\sim\alpha\sim\Lambda\sim$1 within one order of magnitude. In order to understand the origin of this specificity, 
the coupling constant Eq. (\ref{eq:alpha}) can be rewritten with the mass of the mediator m$_0$c$^2\simeq\hbar$c/$a$:

\begin{equation}
\alpha=\frac{1}{\Lambda}\frac{m_0}{m_N}
\label{alcc4}
\end{equation}

showing that the mediator mass is of the same order of magnitude than
the constituent particle mass (m$_p$/m$_\pi$ $\simeq$7 at most): the m$_{\pi,\sigma}\lesssim $m$_N$ fact, responsible for $\alpha \sim$ 1 is embedded in the QCD
realm which provides a typical 1 GeV energy scale, allowing for
similar masses between the lightest baryons and mesons. The present analysis
exhibits the relationship between QCD driven masses ( $\alpha \sim$ 1, Eq. (\ref{alcc4})), the strong LS effect in nuclei ($\eta \sim$ 1) and their QL behavior ($\Lambda \sim$ 1).

In the case of electrons in atoms, which also behave as a QL \cite{mot96} ($\Lambda\sim$1), the
coupling constant is the electromagnetic one: $\alpha$=1/137, yielding $\eta\sim$10$^4$ with Eq. (\ref{keyd2}). Indeed the definition of the spin-orbit parameter (Eq. (\ref{eta})) yields
$\eta\sim$m$_e$c$^2$/V'$_0\sim$ (10$^5$ eV/10 eV) $\sim$10$^4$, which is in agreement with the experimentally observed magnitude of the spin-orbit effect compared to the shell one \cite{ebr15}: 1/$\eta \sim$10$^{-4}$. This result validates the present approach as well as the quantum
liquid nature of electrons in atoms (see Table \ref{tab:quanta}).

In summary three fundamental dimensionless ratios characterise a many-body system, namely the coupling constant, the inertia parameter and the quantality. 
The investigation of the relation (\ref{key}) among these ratios allows to relate interaction, spin-orbit and quantum liquid effects in finite systems. A generalisation
of this relation will discussed in section IV.C.

\subsection{Dispersion and lengthscales}

Localisation properties in nuclei and fermionic many-body systems shall
benefit from the study on dimensionless ratios depicted in the previous section.
In order to characterise the dispersion of constituents, relevant lengthscales 
should be first discussed. On this purpose,
the interconstituent
interaction is approximated by the schematic potential of Fig.
\ref{fig:potint}, as in Appendix of \cite{ebr14}. It should be noted that similar schematic potentials
have been successfully used to investigate the main properties of nuclear
saturation \cite{bm69,gom58}.  This type of approach is used here to
investigate localisation, clusterisation and saturation in a unified
framework.

\begin{figure}[tb]
\begin{center}
\scalebox{0.35}{\includegraphics{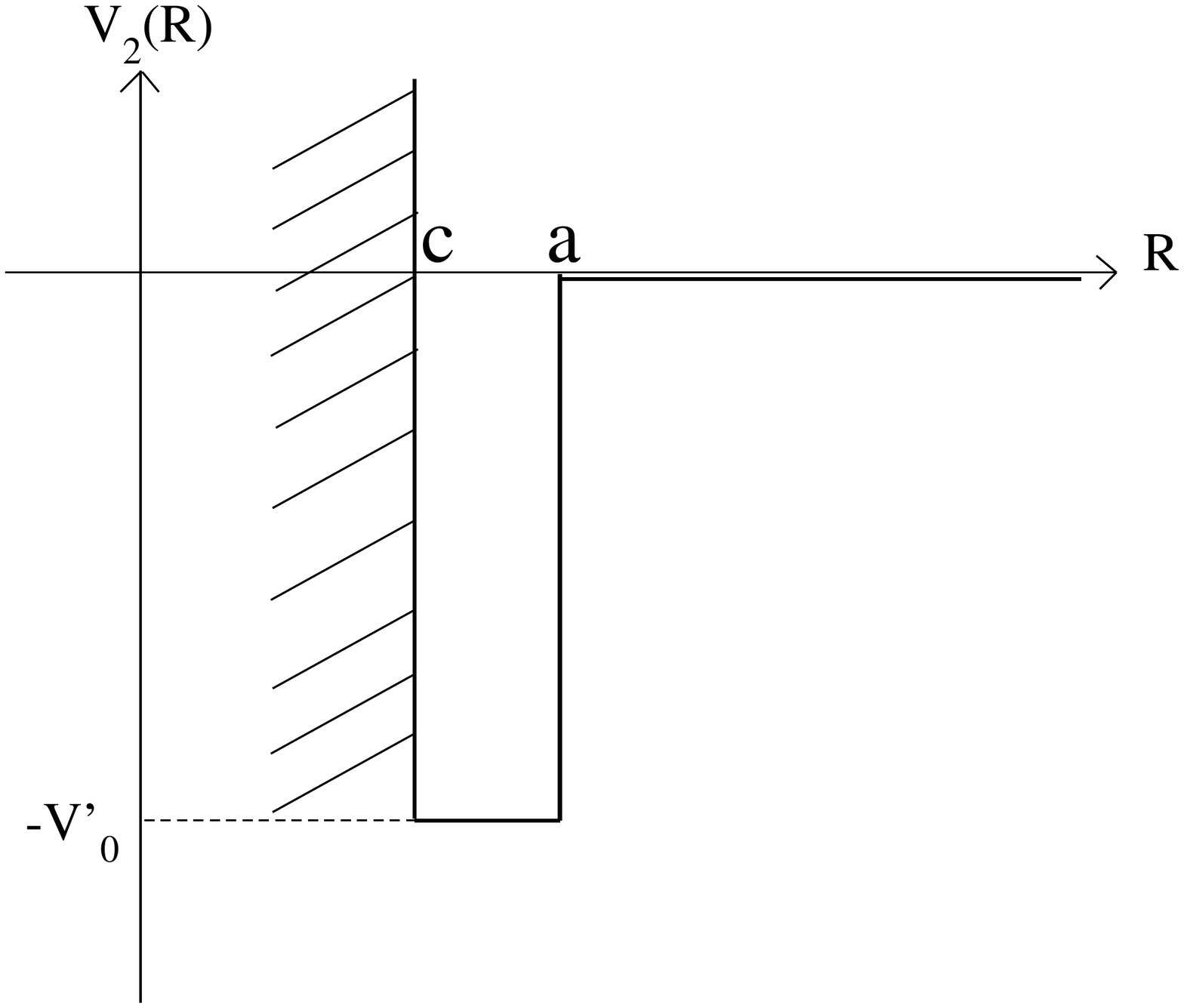}}
\caption{Approximation of the interconstituent potential. c is the hard-core size and a the equilibrium range of the interaction}
\label{fig:potint}
\end{center}
\end{figure}

The derivation in the Appendix of Ref. \cite{ebr14} using the present notations leads to 

\begin{equation}
V_0=\gamma V'_0
\label{eq:v0v0}
\end{equation}

with 

\begin{equation}
\gamma \equiv 1-\left(\frac{c}{a}\right)^3
\label{eq:gam}
\end{equation}

where c and $a$ are the hard-core and the typical range of the interaction (Fig. \ref{fig:potint}), respectively. V'$_0$ it the magnitude of the interaction and V$_0$ the depth of the confining mean-field potential. In nuclei, typical values of $a$ $\sim$ 0.9 fm and
c $\sim$ 0.5 fm provide $\gamma\sim$ 0.8. This is in agreement with the empirical values of V$_0\sim$75 MeV and 
V'$_0\sim$ 100 MeV obtained with Relativistic Mean Field approaches \cite{ebr12,ebr13,ebr14}, showing the validity of this simple model.  

Typical lengthscales can be considered in the light of the present approach, allowing to express the dispersion of the wave function of the constituent from these lengthscales. It has been shown that the oscillator length b characterizes to a good approximation the dispersion of the wavefunctions (see Eq. (\ref{eq:alpha}) and Ref. \cite{ebr13}). Looking for relevant lengthscales within this parameter is therefore in order:

\begin{equation}
b\equiv\frac{\sqrt{\hbar R}}{\left(2m_NV_0\right)^{1/4}}=\sqrt{r_lR}
\label{bsqr}
\end{equation}

where R is the size of the system and

\begin{equation}
r_l\equiv\frac{\hbar}{\sqrt{2m_NV_0}}
\label{eq:defrl}
\end{equation}

The typical dispersion of the constituent is therefore the geometrical average of the total size R of the system with $r_l$. In order to interpret $r_l$,
Eqs. (\ref{eta}) , (\ref{eq:v0v0}) and (\ref{eq:defrl})  can be used:

\begin{equation}
r_l=\sqrt{\frac{\eta}{2\gamma}}.r_N
\label{rlg}
\end{equation}

where $r_N$ is the reduced Compton wavelength (see Appendix A): $r_N$=4.10$^{-3} $\AA\hspace{0.3mm} 
 for electrons and 0.2 fm for nucleons. In the case of electrons in atoms
using either the expression of the binding potential (Rydberg energy) \cite{coh}:

\begin{equation}
V_0=\frac{me^4}{2(4\pi\epsilon_0)^2\hbar^2}
\end{equation}

or the typical value of $\eta \sim $5.10$^4$ (Table I), both lead to the Bohr radius : $r_l=a_0$. $r_l$ is therefore a generalisation of the Bohr radius, which can now be applied to 
other systems than electrons in atoms. In the case of nucleons in nuclei, using the relevant values of $\eta$ and $\gamma$ leads to $r_l \sim $ 0.5 fm which is typically the size of the hard-core of the interaction, or the attractive length of the interaction. 

The present approach on localisation can be generalised to electrons in atoms: Eq. (\ref{bsqr}) shows that the dispersion of an electron is b $\sim \sqrt{a_0.a_ 0}$=$a_0$, in agreement with the analytical expression of the wavefunction of electrons in atoms \cite{coh}. In the case of nucleons in nuclei, their typical dispersion is the geometrical average between the size of the hard-core of the interaction and the total size of the system (Eq. (\ref{bsqr})).

To provide a further interpretation of these lengthscales, the localisation parameter (\ref{eq:alpha}) allows to relate the generalised Bohr radius r$_l$ to the interconsituent distance r$_0$:

\begin{equation}
\alpha_{loc}^2=\frac{R_l}{r_0}
\label{eq:rlr0}
\end{equation}

with $R_l \equiv r_l A^{1/3}$.

\begin{figure}[]  
      {\includegraphics[width=0.50\textwidth]{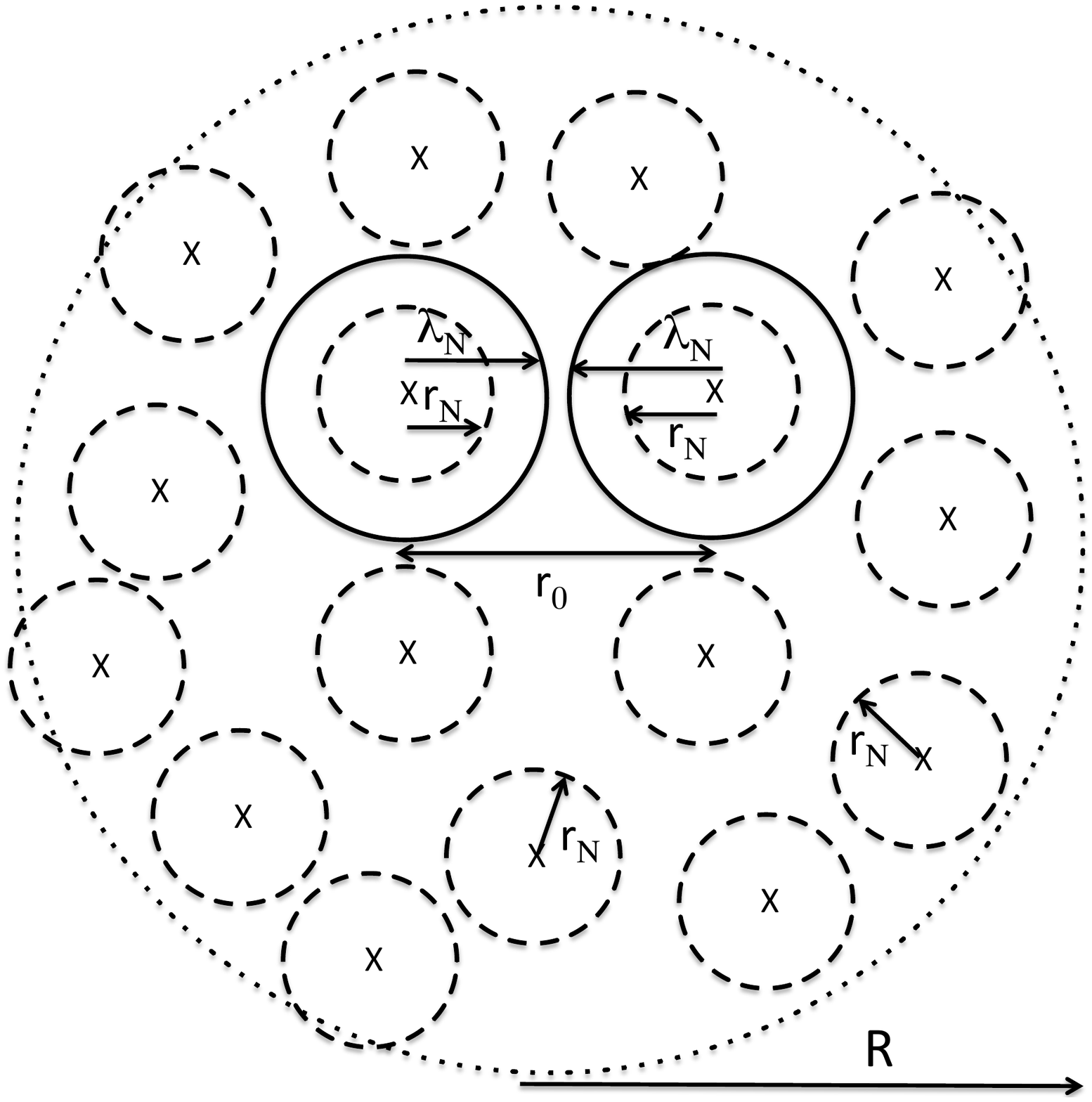}\\
      \includegraphics[width=0.50\textwidth]{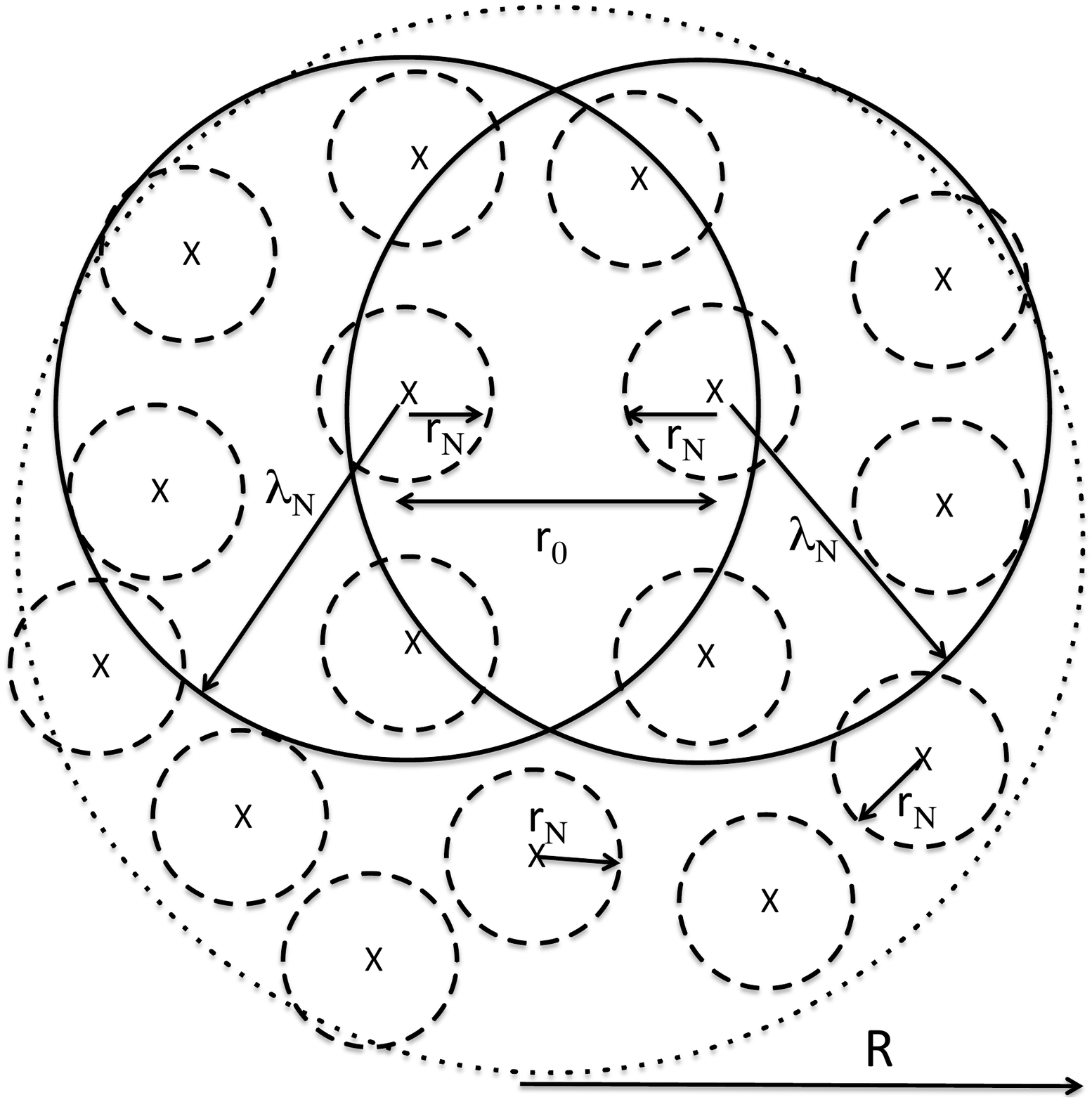}
    }
    \caption{Sketch of the various lengthes, in the
    case of a localised constituent manybody system (top) and in a
    delocalised (nucleus case) one (bottom). r$_N$, r$_0$, $\lambda_N$ and R are the
    constituent reduced Compton wavelength, the constituent
    interdistance, the constituent wavelength and the size of the
    system, respectively. The constituent wavelength $\lambda_N$ 
    is displayed for 2 constituents only.
    The typical spreading of the
    constituent's wave function follows b $\sim \lambda_N$. In nuclei $r_l \sim r_N$, which is close to the hardcore size of the interaction (see Eq. (\ref{rlg}) and Appendix A).}    
    \label{fig:len}
  \end{figure}

In the case of nucleons in nuclei, $R_l$ would correspond to the size of a fully packed system, with all nucleons in hard-core to hard-core contact from each other (because r$_l \sim$ 0.5 fm). Therefore
Eq. (\ref{eq:rlr0}) shows that the system can clusterise ($\alpha_{loc} \sim 1$) when the inter-nucleon distance is large enough to be of the order of the size of the fully packed system. Figure \ref{fig:len} summarizes the various lengthes at work in a manybody system, based
on the present study.

\section{Saturation and clusterisation}\label{sec:ms}

In nuclei the balance between the hardcore plus attractive interaction (localisation effect) and the
QL nature of the system (delocalisation effect) shall allow to describe both the  magnitude of the the saturation of the density \cite{bm69} and the emergence of
cluster states \cite{ebr12,elh16}. 

\subsection{Saturation and delocalisation}

Saturation in nuclei is a fundamental effect, which is described by the hard-core of the interaction as well as QCD rooted effects \cite{weis}. However, as already discussed by Mottelson, the nuclear saturation density
$\rho_0\simeq$0.16 fm $^{-3}$ is somewhat smaller than the packed density, which would correspond to a fully closed packed system of nucleons in contact:
considering the equilibrium range $a$ $\sim$ 0.9 fm (or the hard-core size c $\sim$ 0.5 fm) instead of the interconstituent distance r$_0$ $\sim$ 1.2 fm, leads to a packing density
between a few times and one order of magnitude larger than the saturation density. Hence the lower value of the saturation density (i.e. r$_0>$ a) shall be due to delocalisation effects which remain to be clarified \cite{bm69}. 

The localisation parameter (\ref{eq:alpha}) allows to clear up this point: using Eqs (\ref{la}) and (\ref{eq:v0v0}) the ratio of the saturation density $\rho_0$ to the packed one $\rho_a$ using the equilibrium range $a$ of the interaction, is

\begin{equation}
\frac{\rho_0}{\rho_a}=\left(\frac{a}{r_0}\right)^3=\frac{\alpha_{loc}^6}{A}\left(\frac{2\gamma}{\Lambda}\right)^{3/2}
\label{paro}
\end{equation}

When the quantality $\Lambda$ decreases, the system behaves more like a crystal (Table I), and Eq. (\ref{paro}) shows that the saturation density becomes larger with respect to the packing density, as discussed
in Ref. \cite{bm69}. When the interconstituent distance r$_0$ decreases, the system gets more delocalised: the localisation parameter increases as well as the saturation density of the system. 

Eq. (\ref{paro}) also provides a clarification of the relation between the localisation parameter and the quantality:

\begin{equation}
\alpha_{loc}^4=\frac{\Lambda}{2\gamma}.A^{2/3}.\left(\frac{\rho_0}{\rho_a}\right)^{2/3}
\label{lamalt}
\end{equation}

Therefore, if the interconstituent distance is kept constant, the localisation parameter carries the quantality information in the finite system: a larger value means more delocalisation and a state towards the QL as discussed in Refs. \cite{ebr12,ebr13,ebr14a}.

Apart from the discussion of the magnitude of the saturation density, the $a$ $\sim$ r$_0$ approximation can now be further undertaken in order to discuss delocalisation effects based on two main lengthscales: i) the dispersion of the constituent wavefunction and ii) the interconstituent distance, close to the equilibrium range of the interaction. Such an approximation will 
be made in the following of this work and is valid unless the magnitude of the saturation density is discussed as above.

Using this approximation, Eq. (\ref{lamalt})  yields a simpler relation between the localisation parameter and the quantality:
\begin{equation}
\alpha_{loc}^4=\frac{\Lambda}{2\gamma}.A^{2/3}
\label{alpla}
\end{equation}

where the A$^{2/3}$ factor exhibits the finite size effect of the
localisation parameter. Considering clusterisation, it is relevant to derive the typical number A$_0$ of constituents of the
finite size system for which the hybrid cluster phase occurs : $\alpha_{loc}\simeq$ 1 and
the dispersion of a
constituent particle is of the same order than the interconstituent
distance \cite{ebr12,ebr13}. Eq. (\ref{alpla}) gives in this case

\begin{equation}
A_0\simeq\left(2\gamma/\Lambda\right)^{3/2}
\label{a}
\end{equation}

Eq. (\ref{a}) shows that the number of constituents for which the
cluster phase occurs depends on the quantality. It is of the order
of few units to a couple of dozens in the case of QL ($\Lambda \sim$
0.1, see Table \ref{tab:quanta}) and of the order of 10$^{4-5}$ for a crystal ($\Lambda \sim$
10$^{-3}$), in order to reach the cluster hybrid phase.

\subsection{Quantum liquid to cluster transition in low-density nuclear systems}

When the nuclear density decreases, a transition towards a cluster phase is expected \cite{rop}, such as in the case of an expanding nuclear system \cite{sch13,ebr14a}.
In order to study this effect with the present approach, it is necessary to release the fixed density assumption while maintaining the interconstituent distance r$_0$ constant, as a constraint imposed by the interaction. Hence the localisation parameter (Eq. (\ref{eq:alpha})) shall be generalised as a function of the density  $\rho=A/V$ of the system:
in the case of clusterisation effects related to density, considering for instance an expanding system, the density departs from the saturation density. Returning to the very definition of the localisation parameter and its approximation in the HO case, leads to
 
 \begin{equation}
\alpha_{loc}\equiv\frac{b}{r_0}=\left(\frac{A\rho_0^2}{\rho_l\rho}\right)^{1/6}
\label{eq:almot}
\end{equation}

with $\rho_l \equiv$ 3/(4$\pi r_l^3)$, and r$_l$ the generalised Bohr radius (\ref{eq:defrl}). As discussed in section II.C, r$_l$ is close to the hard-core size of the interaction in the case of nuclei. Therefore in this case $\rho_l$ corresponds to the
the packing density:  using V$_0\simeq$ 80 MeV \cite{ebr12} in Eq. (\ref{eq:defrl}), leads to $\rho_{l}\simeq$1.8 fm$^{-3}$.

Eq (\ref{eq:almot}) is more general than (\ref{eq:alpha}), which is recovered in the case of a system at saturation density ($\rho=\rho_0$). It therefore allows to study the behavior of the system when its density departs from the saturation one as in the case of an expanding nuclear system \cite{rop,sch13,ebr14a}. When the density decreases, the system will eventually clusterise into alpha particles (Mott-like phase transition). This shall corresponds to a state where the sub-systems composed of alpha particles (4 constituents) reach 
$\alpha_{loc}$=1: such an optimal overlap between the four nucleons of the alpha particle generates well defined alpha clusters. In order to study the cluster phase transition in low-density nuclear systems, Eq (\ref{eq:almot}) is inverted:

\begin{equation}
\frac{\rho}{\rho_0}=\left(\frac{A}{\alpha^6_{loc}}\right).\frac{\rho_0}{\rho_l}
\label{eq:mott}
\end{equation}

\begin{figure}[tb]
\begin{center}
\scalebox{0.35}{\includegraphics{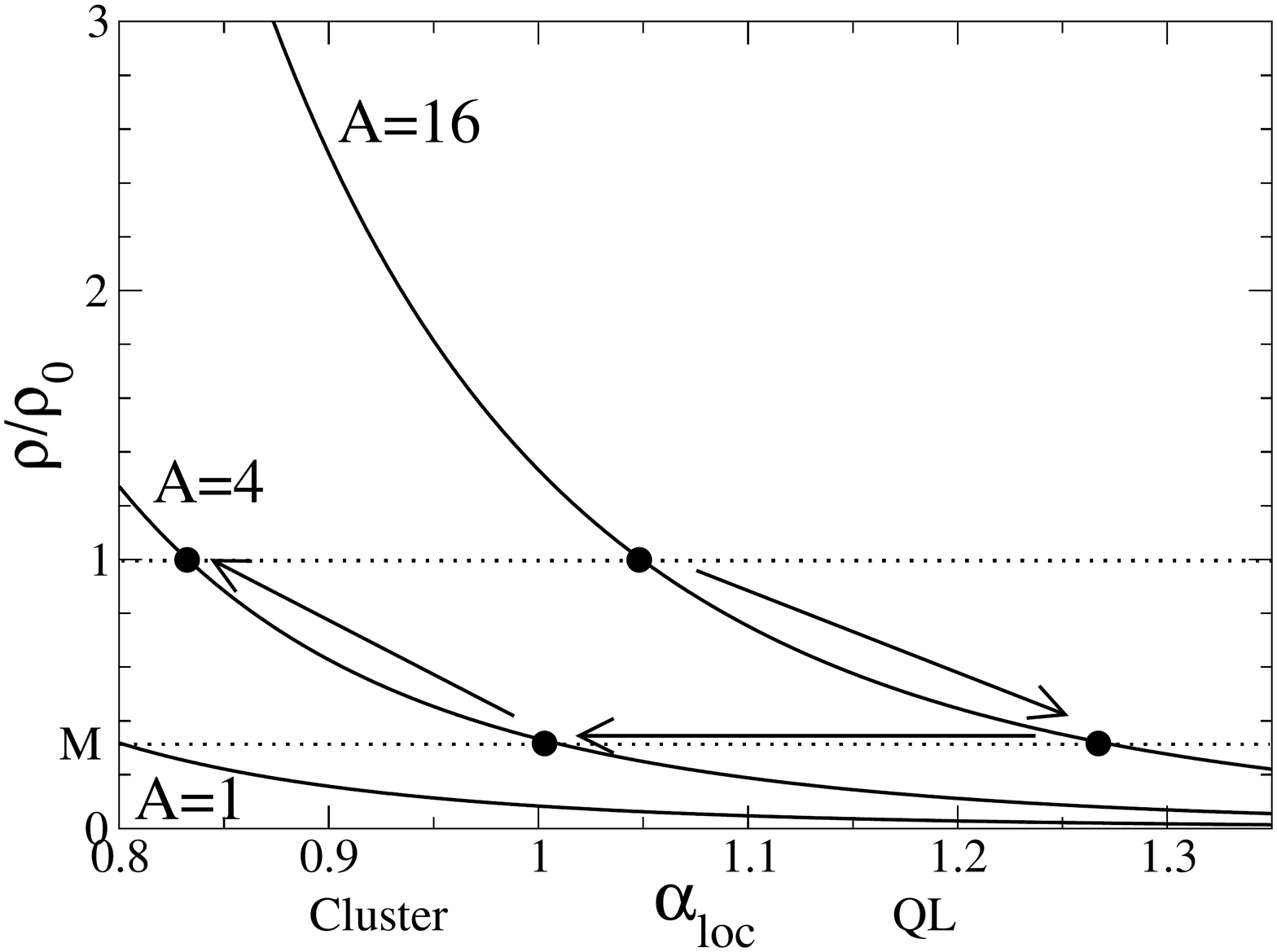}}
\caption{Relative density to the saturation one as a function of the localisation parameter for a systems with 1, 4 and 16 constituents. The two dotted lines correspond to the saturation and Mott-like densities.}
\label{fig:mott}
\end{center}
\end{figure}

Figure \ref{fig:mott} displays the corresponding behavior of the density for the A=1,4 and 16 cases. The Mott-like density $\rho_M$ at which the cluster transition occurs, shall correspond to the $\alpha_{loc}$=1 point on the A=4 curve, as discussed above. This leads to $\rho_M\simeq\rho_0$/3 in striking agreement with several microscopic studies (see \cite{sch13,ebr14a} and Refs therein). The present results show that the cluster phase transition does occur when the alphas reach optimal condition for clusterisation. 
An expanding A=16 system follows the arrows on Figure \ref{fig:mott}: when the density is low enough to reach the Mott-like density, each alpha behaves as a cluster. The alphas then reach back saturation density due to the nuclear interaction favoring this density. The present approach is therefore in both qualitative and quantitative agreement with more microscopic studies of Refs. \cite{ebr14a,sch13}.

In order to investigate this effect in more details, two specific cases are considered: system at saturation or at Mott-like density. In nuclei Eq (\ref{eq:mott}) leads to

\begin{equation}
\alpha_0\equiv\alpha_{loc}({\rho_0})\simeq \frac{2}{3}A^{1/6}
\label{eq:alsat}
\end{equation}

and

\begin{equation}
\alpha_M\equiv\alpha_{loc}({\rho_M})\simeq 0.8 A^{1/6}\simeq 1.2 \alpha_0
\label{eq:almot2}
\end{equation}

The $\alpha_0$ case corresponds to Eq (1), namely the density of the system constrained to the saturation one. The occurrence of cluster or QL states in this specific case has been previously studied \cite{ebr12,ebr13,ebr14a}: $\alpha_0$=1 leads to A$\simeq$10 (Eq (\ref{eq:alsat})), showing that clusterisation is favored in light nuclei as discussed in Ref. \cite{ebr13}.
More generally, the density $\rho_C$ for which clusters are favored ($\alpha_{loc}$=1) is, from Eq (\ref{eq:almot}):

\begin{equation}
\frac{\rho_C}{\rho_0}=A\frac{\rho_0}{\rho_l}\simeq 0.08 A
\label{eq:clusr}
\end{equation}
 
Therefore light (A$\sim$10) nuclei favor clusterisation (i.e. $\rho_C \simeq \rho_0$ for A $\sim$ 10) because the ratio of the packing density to the saturation one is about 10.

Eq (\ref{eq:almot}) shows that $\alpha_M$=1 is reached for A=4, whereas A=2 or A=6 depart from this value. The present approach therefore describes in a unified way that clusters states are likely in light nuclei, and also that systems undergo a transition to alpha-clusterisation when the density decreases to about 1/3 of the saturation density, providing a complementary light on this low-density phase transition.

\subsection{Mean-field and spatial dispersion}

The impact of the interconstituent interaction on delocalisation effects can be investigated through the mean-field at work in the system: the confining mean-field is a relevant tool to relate the nucleonic interaction, localisation effects and eventually nuclear saturation.
Providing an explicit form of the attractive and repulsive mean-field potentials at work in the system allow to bridge these quantities.

 In this framework meson exchange studies in a relativistic scheme provide an appropriate ground to link the nucleonic interaction to the mean-field potential \cite{wal}. 
In a simple approach, the attractive and repulsive parts are described by the propagation of so-called $\sigma$ and $\omega$ meson, respectively, with m$_\omega >$ m$_\sigma$. At the mean-field level, the $\sigma$ meson generates an attractive scalar S potential and the $\omega$ meson a repulsive vector V one. The total magnitude of the confining potential is therefore V$_0$= -V-S.

Relating the meson exchange picture to the features of the nucleonic interaction (Fig. \ref{fig:potint}), the Yukawa lengths correspond to $\mu^{-1}_\sigma\simeq$ $a$ and $\mu^{-1}_\omega\simeq$ c, respectively. Defining the relative difference of the meson masses:
\begin{equation}
\nu \equiv \frac{m_\omega - m_\sigma}{m_\omega}=1-\frac{c}{a} < 1
\label{eq:nudeft}
\end{equation}

provides with Eq. (\ref{eq:gam}):
\begin{equation}
\gamma=(\nu -1)^3 +1
\label{eq:gnu}
\end{equation}

with 0 $< \nu <$ 1.

Considering the potentials at distance c and assuming similar magnitudes of the coupling constants for the $\omega$
and $\sigma$ Yukawa potentials, gives with (\ref{eq:v0v0}): 

\begin{equation}
S=-\frac{\gamma V'_0}{1-e^{-\nu}}
\label{eq:snu}
\end{equation}

and 

\begin{equation}
V=\frac{\gamma V'_0}{e^\nu -1}
\label{eq:vnu}
\end{equation}

Let us apply the present formalism to the case of nuclei. In the Relativistic Mean-Field (RMF) approach, $m_\sigma c^2$=550 MeV and $m_\omega c^2$=782 MeV (which corresponds to their bare masses), leading to $\nu\simeq$0.3 (Eq. (\ref{eq:nudeft})). A typical energy scale V'$_0$=100 MeV of the magnitude of the nucleon-nucleon interaction provides V$_0\simeq$ 70 MeV, S$\simeq$ -250 MeV and V$\simeq$ 200 MeV through Eqs. (\ref{eq:v0v0}), (\ref{eq:gnu}), (\ref{eq:snu})  and  (\ref{eq:vnu}). Considering the simplicity of the model, these values are in qualitative agreement with the typical magnitudes obtained from a microscopic RMF calculation \cite{vre05}: S=-400 MeV and V=350 MeV (Eqs. (\ref{eq:snu}) and (\ref{eq:vnu})). It should be noted that Eqs (\ref{eq:snu}) and (\ref{eq:vnu}) yield S=-Ve$^\nu$, with $\nu\rightarrow$ 0 recovering the pseudo-spin symmetry limit (same masses for the attractive and repulsive mesons).   

\begin{figure}[tb]
\begin{center}
\scalebox{0.35}{\includegraphics{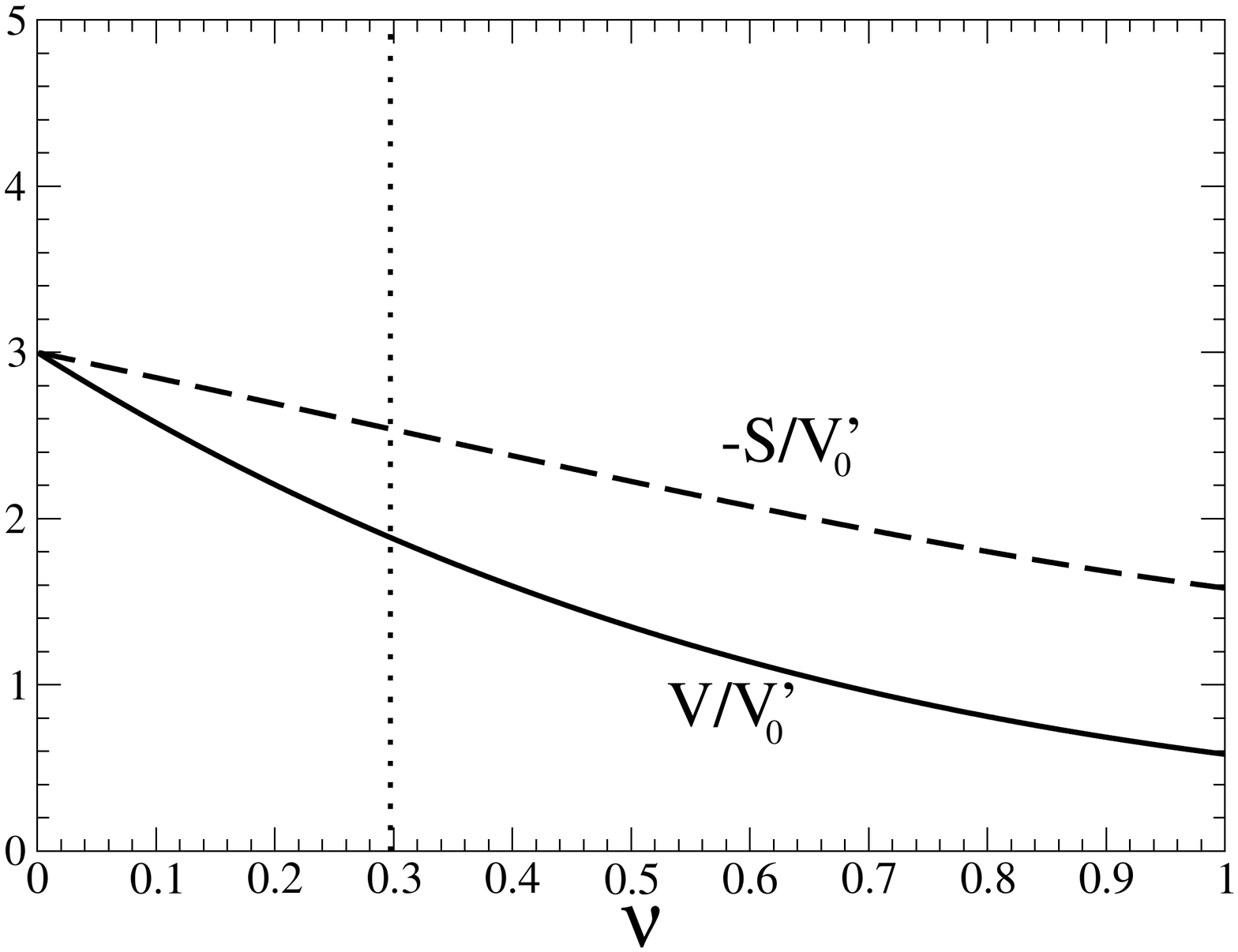}}
\caption{Scalar and vector potentials as a function of the relative difference of mesons masses.
The vertical dotted line corresponds to the case of nuclei.}
\label{fig:potvs}
\end{center}
\end{figure}

The above expressions show that the mean potentials at work in the system are driven by two quantities: the typical magnitude of the interconstituent interaction V'$_0$ (which introduces the energy scale), and  the dimensionless quantity $\nu$ which is the relative difference of masses of the mesons. The corresponding behaviors of the vector and scalar potentials are displayed on Fig. \ref{fig:potvs}. Both V and S remain within a few times the value of the typical magnitude V'$_0$ of the interconstituent interaction due to the hard-core plus attractive behavior of the interaction.

Let us now investigate the respective roles of both the delocalisation and the hard-core plus attractive feature of the nucleonic interaction. Eqs (\ref{bsqr}) and (\ref{rlg})
show that the dispersion of the wave function behaves as:

\begin{equation}
b=\sqrt{r_NR}\left(\frac{\eta}{2\gamma}\right)^{1/4}
\label{betar}
\end{equation}

Both pivotal contributions of the nucleonic interaction and the delocalisation effects to the mean-field potentials (which generates saturation) are encoded in the present approach through Eqs. (\ref{eq:gnu}) and (\ref{betar}): a large delocalisation is obtained for a small $\gamma$ (interaction effect), that is a small $\nu$. A small $\nu$ value corresponds to similar mesons masses, or a small attractive width: the delocalisation behavior is partly rooted in the balance between the attractive and repulsive parts of the interaction, i.e. 0 $\lesssim\nu\lesssim$  1. It should be noted that in such systems, the V and -S values shall be about 3 times the magnitude of the interconstituent interaction as displayed on Fig. \ref{fig:potvs} for small $\nu$ values. This feature shall be now used to discuss the spin-orbit effect.

\section{Properties of many-body systems}

The present approach can be used to study properties of finite many-body systems such as the spin-orbit effect. A general picture of localisation in such systems can also be considered. Finally, additional relevant dimensionless quantities shall be derived.

\subsection{Spin-orbit coupling in finite systems}\label{sec:ls}

\subsubsection{Discussion on the inertia parameter}

The inertia parameter (\ref{eta}) allows to understand the behavior of the spin-orbit effect in various many-body systems, leading to the so called LS rule \cite{ebr15}. 
It should be noted that the V-S quantity is considered in the denominator of the inertia parameter in the case of the LS rule,  instead of V'$_0$.
Both the vector V and scalar S potentials described in the previous section are of the same order of magnitude than V'$_0$, as seen on Fig. \ref{fig:potvs}: more precisely, V'$_0$ $\sim$ 100 MeV and 
V-S $\sim$ 750 MeV \cite{ebr12} leading to 1.2 $<$ $\eta$ $<$ 9.4 depending if either V'$_0$ or V-S
is considered in the denominator of Eq. (\ref{eta}). $\eta$ therefore remains of a few units in both cases and both $\eta \equiv$ m$_N$/V'$_0$ and $\eta \equiv$ m$_N$/(V-S) are relevant for a qualitative discussion.

The simple model described in the previous section provides an explanation for the typical factor 7 between 
V-S and V'$_0$. The quantity of interest here is the (V-S)/V'$_0$ ratio, in order to investigate the factor impacting the denominator of $\eta$: $\eta=m_N/V'_0$, as discussed in 
section II or more precisely $\eta=m_N/(V-S)$ when using a Schr\"odinger-like reduction of the Dirac equation \cite{ebr15}.  Eqs (\ref{eq:snu}) and (\ref{eq:vnu}) give:

\begin{equation}
\frac{V-S}{V'_0}=\left[(\nu -1)^3 +1\right ] coth\left(\frac{\nu}{2}\right)
\label{eq:vms}
\end{equation}

\begin{figure}[tb]
\begin{center}
\scalebox{0.35}{\includegraphics{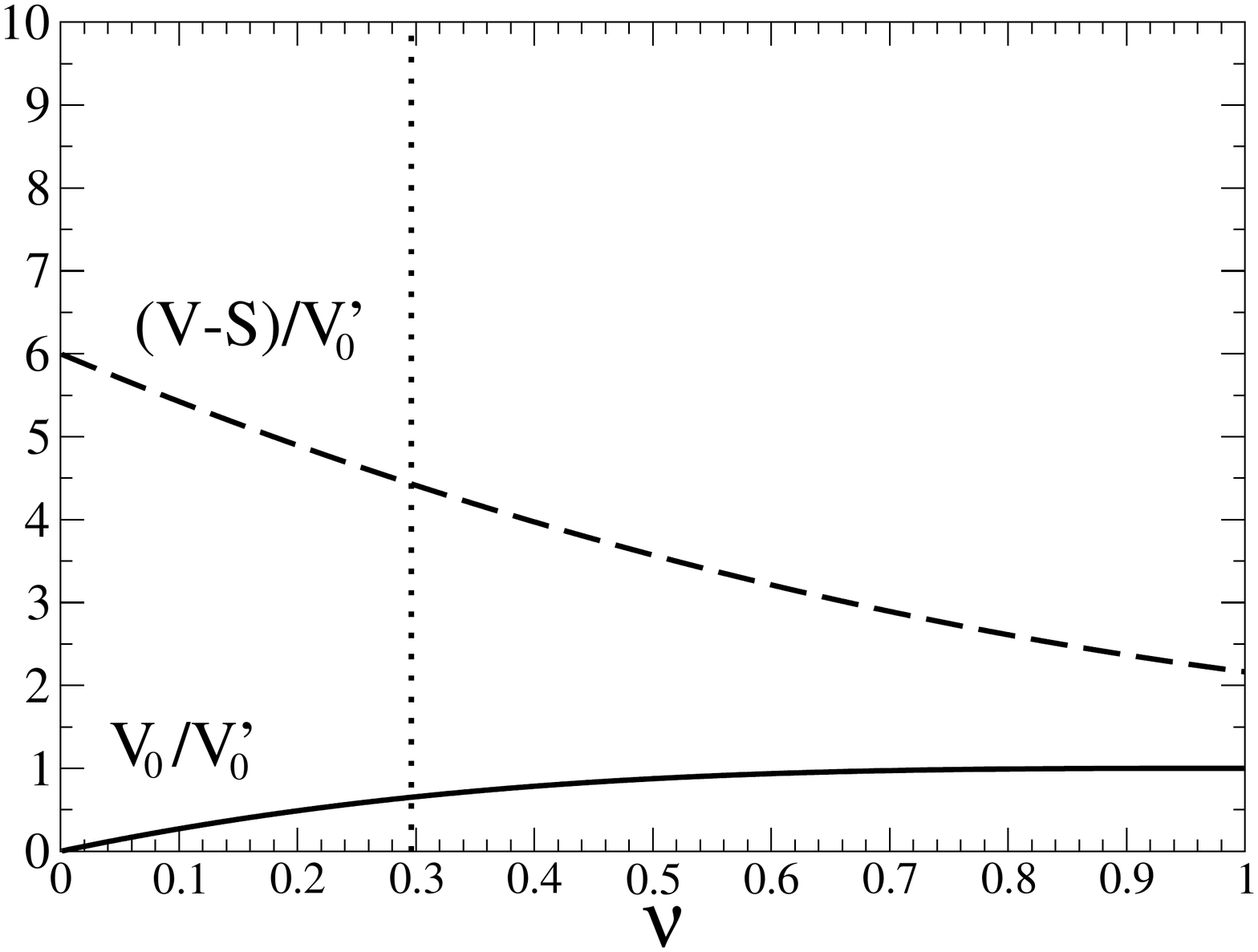}}
\caption{The mean-field confining potential (solid line) and the one driving the spin-orbit effect (dashed lines) as a function of the relative difference of the 
meson masses. The vertical dotted line corresponds to the case of nuclei.}
\label{fig:potls}
\end{center}
\end{figure}

The behavior of both the mean confining potential V$_0$ (Eq. (\ref{eq:v0v0})) and the V-S potential (Eq. (\ref{eq:vms})) are displayed on Fig. \ref{fig:potls}:
the V-S potential, driving the spin-orbit effect, remains between 2 and 6 times the magnitude of the interconstituent interaction and does not
depend on the mass of the constituent. In the case of a small relative difference of masses of the mesons ($\nu\rightarrow$0), a first order expansion of  Eq.(\ref{eq:vms}) leads to (V-S)/V'$_0\simeq$ 6(1-$\nu$): the factor 6 for V -S compared to V'$_0$ automatically comes out from the small mass difference between the two mesons, that is similar order of magnitude of lengthscales between the attractive and repulsive parts of the interconstituent interaction. The present model, despite its simplicity, nicely describes the factor (V-S)/V'$_0\sim$7 discussed above. 

\subsubsection{The spin-orbit rule}

The spin-orbit effect is known to play a relevant role in finite quantum liquids such as nuclei and electron in atoms. It's magnitude is much larger in the former than in the latter.
In the case of crystals, the spin-orbit effect becomes even much smaller, typically by a factor 10$^{-3}$ due to the intermolecular term, compared to the intramolecular one \cite{coo78}.  In this subsection, phenomenological aspects related to the spin-orbit effect are investigated using the present approach.

In the case of finite quantum liquids, the Dirac equation of motion brings an additional relation between V'$_0$,
r$_0$ and m$_N$ \cite{ebr15}. Considering the relative
magnitude $x$ of the HO to the LS gap:

\begin{equation}
\label{x_ratio}
x\equiv\frac{\hbar\omega_0}{|\Delta < V^{LS}>|}
\end{equation}

the spin-orbit rule allows to relate the inertia parameter $\eta$ to 
$x$: the use of the Dirac equation provides  \cite{ebr15}

\begin{equation}
x\simeq \bigg| \eta-1+\frac{1}{4\eta} \bigg|
\label{x2}
\end{equation}

with $\eta$=(V-S) as discussed in the previous subsection.
 
\begin{figure}[tb]
\begin{center}
\scalebox{0.35}{\includegraphics{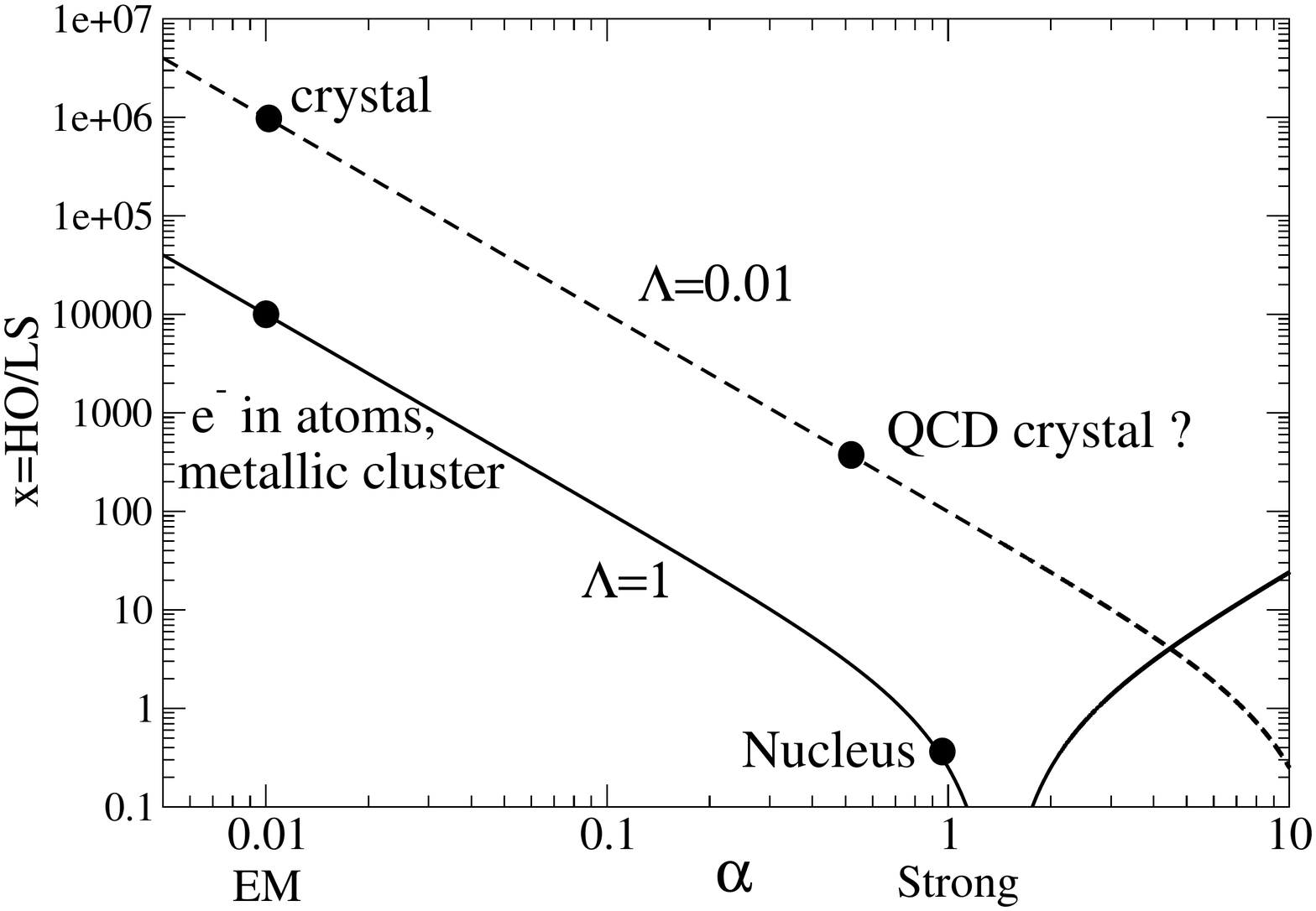}}
 \caption{The relative magnitude of the HO to LS gaps in a many body system as a function of the dimensionless coupling constant with $\Lambda$=1 (QL, solide line) 
 and $\Lambda$=0.01 (crystal, dashed lines) (Eq. (\ref{xalpha})).}    
\label{fig:spcou}
\end{center}
\end{figure}

Using the key relation (\ref{key}) Fig. \ref{fig:spcou} displays the $x$($\alpha$) relation, for
both the QL and crystal cases:

\begin{equation}
x(\alpha)=\frac{1}{\Lambda\alpha^2}-1+\frac{\Lambda\alpha^2}{4}
\label{xalpha}
\end{equation}

The $x(\alpha)$ relation (\ref{xalpha}) generalises the concept
of {\it fine structure} constant: in the specific case of a quantum liquid ($\Lambda\simeq $1) 
and in the case of the electromagnetic interaction ($\alpha\simeq$10$^{-2}$),
Eq. (\ref{xalpha}) reduces to

\begin{equation}
x(\alpha)\simeq\frac{1}{\alpha^2}
\end{equation}

recovering the specific fact that the coupling constant is driving the magnitude of the spin-orbit effect of electron in atoms \cite{som}. 
Eq. (\ref{xalpha}) therefore provides a more general understanding of the impact of the coupling constant on the
spin-orbit effect, in various systems.

Figure \ref{fig:spcou} shows that in a QL ($\Lambda\simeq $1), the large or giant spin-orbit coupling (x$<<$1)
can only be reached in strongly interacting systems whereas the small
spin-orbit coupling (x$>>$1) can only be reached in electromagnetic interacting
systems:  a large spin-orbit effect requires both a coupling constant of the order of unity (such as in the strong
interaction case) and a quantum liquid behavior as seen on Eq. (\ref{xalpha}). 

In the case of a
crystal ($\Lambda <<$1), the whole curve is shifted to larger x values. The LS effect
are therefore expected to be strongly reduced, by a factor corresponding to the drop of
the quantality value from a QL to a crystal, typically 2-3 orders of
magnitude (Table I). This result is in agreement with dedicated calculations of the spin-orbit effect in crystals \cite{coo78}.

\subsection{The generalised localisation}
\label{sec:gll}

As discussed above, Mottelson used the quantality to
analyse when a system behaves like a quantum liquid or a crystal \cite{mot96}.
As we shall show here, this approach is suited for most of the
many-body systems, but could be further generalised. We will here make use of several results on lengthscales derived in Appendix A.

From a general point of view, the localisation property of a system is driven by the
$\lambda_N$/r$_0$ ratio where $\lambda_N$ is the constituent wavelength (see Eq (\ref{dbr})) \cite{ebr12}.
When $\lambda_N$ is larger
than the typical interconstituent distance r$_0$, the system reaches
a QL state. The inverse case corresponds to a crystal state.

Approximating the typical kinetic energy of a constituent to the magnitude of the interaction (i.e. T$_N \sim $ V'$_0$) and using Eqs. (\ref{dbr}) and (\ref{la}) allows to obtain, in the $a$ $\sim$ r$_0$ approximation:

\begin{equation}
\frac{\lambda_N}{r_0}=\pi\sqrt{2\Lambda}
\label{eq:locla}
\end{equation}

which shows that the quantality does drive the localisation property of
the system. This relation between the localisation and the
quantality has been introduced by Mottelson, and analysed in subsequent
works \cite{zin13,ebr14}. 

Using (\ref{lamact}), Eq. (\ref{eq:locla}) becomes

\begin{equation}
\frac{\lambda_N}{r_0}=\frac{\pi\sqrt{2}}{{\cal A}} \simeq
\frac{5}{{\cal A}}
\label{eq:locact}
\end{equation}

If quantum effects are small, then $\cal{A}$$\gg$ 1 and therefore
$\lambda_N$ $\ll $r$_0$, which is expected in the crystal case. If
quantum effects are large, $\cal{A}$$\simeq$1 and $\lambda_N$ 
$\sim$ 5r$_0$ is the order of magnitude of the maximal delocalisation
which can be reached in a QL.

However the derivation of Eq. (\ref{eq:locla})
relies on two assumptions: first, the system shall be non-relativistic and
second, the kinetic energy of the constituent T$_N$ shall be about the
V'$_0$ value: T$_N\simeq$V'$_0$. It is therefore possible to generalise
Eq. (\ref{eq:locla}) to all kind of kinematics, i.e. releasing both the non-relativistic assumption and the 
T$_N\simeq$V'$_0$ one. In this case the use of Eqs. (\ref{cc2}) and
(\ref{co2}) provide the general expression for the localisation:

\begin{equation}
\frac{\lambda_N}{r_0}=\frac{2\pi\alpha\Lambda}{\sqrt{\left(\frac{T_N}{m_Nc^2}\right)^2+
2\frac{T_N}{m_Nc^2}}}
\label{deltop}
\end{equation}

Thus, in its most general form, the localisation properties of the
system not only depends on $\Lambda$ as in Eq. (\ref{eq:locla}), but
also on the coupling constant $\alpha$ of the interaction at work. More precisely, localisation
depends on the product $\Lambda\alpha$ which is the quantum fluidity, defined and discussed in section IV.C.
The value of the quantum fluidity is displayed in Table \ref{tab:quanta} for various systems. 
The localisation Eq. (\ref{deltop}) also depends on the kinematics of the constituent through the
T$_N$/m$_N$c$^2$ ratio: a smaller kinetic energy generates a more spread wave function.

\begin{figure}[tb]
\begin{center}
\scalebox{0.35}{\includegraphics{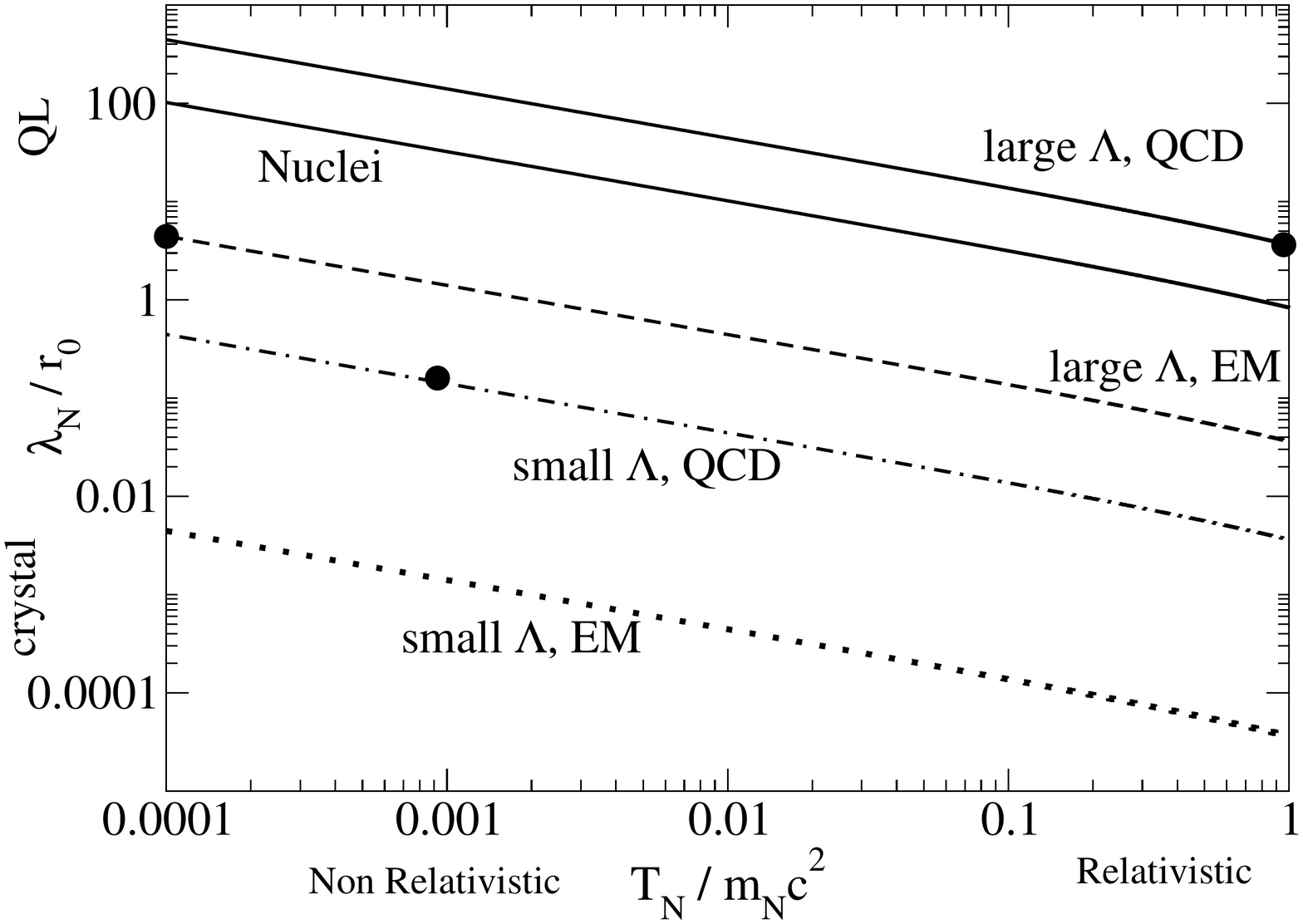}}
 \caption{Localisation parameter (Eq.
(\ref{deltop})) as a function of the ratio of the constituent kinetic
    energy to its mass, for 4 typical cases. Full black dots correspond to the T$_N$=V'$_0$ case.
     Small and large $\Lambda$ correspond to $\Lambda\sim$10$^{-3}$ and $\Lambda\sim$1, respectively (see Table \ref{tab:quanta}). }    
    \label{fig:deloc}
\end{center}
\end{figure}

Fig. \ref{fig:deloc} displays the behavior of the generalised localisation ratio 
(\ref{deltop}) for 4 typical cases, as well as for nuclei. 
As expected, the delocalisation is favored for small kinetic energy of the
constituent, providing a more quantitative description of this effect through Eq. (\ref{deltop}).

In order to investigate the behavior of the generalised localisation, two limit cases are considered: T$_N$=V'$_0$ and
T$_N$=T$_0$ (the zero point kinetic energy).

\subsubsection{The T$_N$=V'$_0$ case }
  
This corresponds to a system where the kinetic energy of the constituents is of the order of the magnitude of their interaction, which is the case of the majority
of the many-body systems. In the specific T$_N$=V'$_0$ case, the generalised localisation (Eq. (\ref{deltop})) becomes:

\begin{equation}
\frac{\lambda_N}{r_0}=\pi\sqrt{2\Lambda}.\left(1+\frac{1}{2\eta}\right)^{-1/2}
\label{delred}
\end{equation}

The dependence on the inertia parameter $\eta$ is related to the fact that 
$\eta$ is a probe for the importance of various relativistic
effects in the system, as discussed in section II.B.2.

 Eq.
(\ref{delred}) is a more general expression than Eq. (\ref{eq:locla}) for the evaluation of the localisation properties of constituents of the system.
The generalised localisation parameter $\lambda_N$/r$_0$ only depends on $\Lambda$ and $\eta$ showing the relevance of the present approach based on the three dimensionless ratios $\Lambda$, $\eta$ and $\alpha$.
Only in the non-relativistic limit, $\eta$ diverges and therefore Eq.
(\ref{delred}) reduces to Eq. (\ref{eq:locla}). This is relevant if the magnitude of the interaction V'$_0$ remains 
small compared to the mass of the constituent m$_Nc^2$.

The present study therefore raises the question if many-body systems with relativistic kinematics 
can be found in nature. It seems that graphene could correspond to this case in an effective way (Appendix B). This may also be the case of electrons in heavy atoms,
which can reach a binding energy of few tens of keV, which is not
negligible compared to their mass. 

The ratio of Eq. (\ref{delred}) to Eq. (\ref{eq:locla}) allows to
evaluate relativistic effects which are not considered in  Eq.
(\ref{eq:locla}), namely

\begin{equation}
\left(1+\frac{1}{2\eta}\right)^{-1/2}=\left(1+\frac{\Lambda\alpha^2}{2}\right)^{-1/2}
\label{delrat}
\end{equation}
 
where the key relation (\ref{key}) has been used. Therefore
$\Lambda\alpha^2$ is the relevant quantity to evaluate relativistic
effects. This is in agreement with the fact that the T$_N$=V'$_0$ condition corresponds to
T$_N$/m$_N$c$^2$=$\eta^{-1}$=$\Lambda\alpha^2$, which measures the relativistic effects.  For instance, in the case of a low energy QCD
interaction with a large $\Lambda$ value ($\eta\sim\Lambda\sim\alpha\sim$1), Eq. (\ref{delrat}) gives 
a $\sqrt{2/3}$ factor. This is the typical impact of relativistic effects
which are neglected in Eq. (\ref{eq:locla}). This effect is therefore small in the case considered here, 
but it may be relevant to investigate more limit cases such as the graphene one as discussed in Appendix B.

The value corresponding to the T$_N$=V'$_0$ limit case is 
displayed on Figure \ref{fig:deloc} using Eq. (\ref{delred}) with typical $\eta^{-1}$=$\Lambda\alpha^2$ value
for each case (full black dots on Fig. \ref{fig:deloc}).

\subsubsection{The T$_N$=T$_0$ case}
\label{subcol}

Eq. (\ref{deltop}) shows that the constituent of any
system can tend towards delocalisation, when its kinetic energy
T$_N$ decreases. For instance on Fig. \ref{fig:deloc} the large $\Lambda$ system with the electromagnetic interaction
can behave either like a QL or a crystal, depending on the T$_N$/m$_N$c$^2$ ratio. This shows that the quantality is not the main quantity driving the localisation properties of a system, in the view of the general expression (\ref{deltop}). One should therefore investigate the regime of low kinetic energy.

The limit case corresponds to the minimal kinetic energy which can be reached by a constituent, namely its zero point kinetic energy T$_0$.
This would correspond to cooled systems. It should be noted that in the case of electrons in metals, V'$_0<$T$_0$ (i.e. $\Lambda > $1, Table I) hence the case 
T$_N$=V'$_0$ cannot be reached in this system.

In the T$_N$=T$_0$ case, Eq. (\ref{deltop}) becomes:

\begin{equation}
\frac{\lambda_N}{r_0}=\pi\sqrt{2}.\left(1+\frac{(\alpha\Lambda)^2}{2}\right)^{-1/2}
\label{delredc}
\end{equation}

The quantum fluidity $\Lambda\alpha$  (section IV.C) is therefore driving this maximal delocalisation value. Since in most systems $\Lambda\alpha <<$ 1 (Table \ref{tab:quanta}), Eq. (\ref{delredc}) shows that $\lambda_N$ $\sim$ 5r$_0$ is the order of magnitude of the maximal delocalisation
which can be reached, as discussed above. It should be noted that the larger the quantum fluidity, the smaller the delocalisation: the constituents have a larger zero point motion (and kinetic energy since T$_N$=T$_0$) and their wavelength get smaller.

In summary the most general way to evaluate the QL vs crystal nature
of a many-body system has been provided with Eq. (\ref{deltop}), showing an essential disentanglement between delocalisation and quantality.
The delocalisation is only driven by the quantality (Eq. \ref{eq:locla}) in the specific T$_N$=V'$_0$ regime, but also depends on $\eta$
when relativistic effects are considered (Eq. \ref{delred}). Looking to the low kinetic energy regime (T$_N$=T$_0$), the delocalisation 
is then rather driven by the quantum fluidity (Eq. \ref{delredc}). 

\subsection{A general approach to dimensionless ratios and their interpretation}

The present study shows the pivotal role of the $\alpha$,$\eta$ and $\Lambda$
ratios to characterize several fundamental properties of many-body states. Emerging quantities such as the quantum fluidity have also been discussed.
The aim of this section is to build in a systematic way dimensionless
quantities, irrespective of their possible physical interpretation,
from the 3 basic quantities of the system: V'$_0$, r$_0$ ($\sim$ $a$) and
m$_N$. This shall allow to produce new dimensionless quantities in the spirit of the quantality and the quantum fluidity, which
may have physical meaning and which may not have been considered so far.

\subsubsection{Systematic derivation of dimensionless ratios}

In order to have only energy units, and for the sake of simplicity, the r$_0$ lengthscale can equivalently be
described by the following mass:

\begin{equation}
m_0c^2 = \frac{\hbar c}{r_0}
\label{yu}
\end{equation}

In the r$_0 \sim$ $a$ approximation, m$_0$ would correspond to the typical mass of the mediator of the interaction.

The most general dimensionless quantity which can be built out of V'$_0$, r$_0$ and m$_N$ is then:

\begin{equation}
\Lambda_{abc}\equiv(m_Nc^2)^a.(m_0c^2)^b.(V'_0)^c
\label{gene}
\end{equation}

with a+b+c=0 (here a, b, c are only the exponents into Eq. (\ref{gene}) and are unrelated 
to physical quantities previously discussed). This condition yields c=-a-b and
therefore $\Lambda_{abc}$=$\Lambda_{ab}$. 

It is convenient to express (\ref{gene}) as a function of two
independent dimensionless ratios of the basic quantities of the
system. We choose the inertia parameter (\ref{eta}):

\begin{equation}
\eta\equiv\frac{m_Nc^2}{V'_0}
\label{etaa}
\end{equation}

and the dimensionless coupling constant (\ref{cc}):

\begin{equation}
\alpha=\frac{V'_0}{m_0c^2}
\label{cca}
\end{equation}

Using (\ref{etaa}) and (\ref{cca}) with (\ref{gene}) yields

\begin{equation}
\Lambda_{ab}=\eta^a.\alpha^{-b}
\label{gene2}
\end{equation}

Eq. (\ref{gene2}) is a generalisation of the key
relation Eq. (\ref{key}). As discussed in section II.B, this shows that
$\eta$ and $\alpha$ form a complete set of quantities able to capture any general
behavior of a manybody system. Any dimensionless quantity built from
V'$_0$, r$_0$ and m$_N$ can be expressed from $\eta$ and $\alpha$. 

It is relevant to further investigate only the a$<$0 case since the a$>$0 case
provides the inverse value of the dimensionless quantity
$\Lambda_{-a-b}$. It should also be noted that fractional values of a, b, or c, can always be
reduced to integer cases, as it is for example between the action and
the quantality: Eq.(\ref{lamact}) translates into 

\begin{equation}
\Lambda_{-1,2}=\Lambda^{-2}_{1/2,-1}
\label{ladeqac}
\end{equation}

and more generally Eq.(\ref{gene}) implies

\begin{equation}
\Lambda^d_{a,b}=\Lambda_{ad,bd}
\end{equation}

In the case of up to maximum second order for any of
the 3 basic quantities of the system (V'$_0$, m$_0$ and m$_N$), the
only dimensionless quantities which can be built are: 

\begin{equation}
\Lambda_{-1,2}=\frac{m_0^2c^4}{m_Nc^2V'_0}
\label{lad1}
\end{equation}

which is the quantality $\Lambda$ (see Eq. (\ref{key})),

\begin{equation}
\Lambda_{-1,-1}=\frac{V'^2_0}{m_0c^2.m_Nc^2}
\label{lad3}
\end{equation}

and

\begin{equation}
\Lambda_{-2,1}=\frac{m_0c^2V'_0}{m_N^2c^4}
\label{lad2}
\end{equation}

The physical interpretation of $\Lambda_{-1,-1}$ and $\Lambda_{-2,1}$
could be further investigated. For instance: 

\begin{equation}
\Lambda_{-1,1}=\frac{1}{\eta\alpha}=\alpha\Lambda=\sqrt{\frac{T_0}{m_Nc^2}}
\label{eqqm}
\end{equation}

where Eqs (\ref{gene2}), (\ref{key}) and (\ref{la}) have been used. $\Lambda_{-1,1}$ is the quantum fluidity discussed in section IV.B, which can impact the localisation properties in many-body system.
Eqs (\ref{eqqm}) is the ratio of the zero point kinetic energy to the mass of the constituent, that is the motion generated by specific quantum effects.
Typical values of the quantum fluidity are given in Table I for various systems. $\Lambda_{-2,1}$ could be interpreted as a quantum mobility effect and is discussed below.

Other quantities such as  $\Lambda_{-1,-2}$, $\Lambda_{-2,-1}$,
$\Lambda_{-1,-1}$, etc. involve at least one of the 3 basic
quantities of the system (V'$_0$, m$_0$ and m$_N$) with a cubic
exponent, which may become more difficult to interpret. However all these
quantities remain dimensionless quantities that could be built from
$\alpha$ and $\eta$, and could be considered for a physical interpretation as in the quantality case.

\subsubsection{The quantum mobility}

In order to interpret the quantity $\Lambda_{-2,1}$ (Eq. (\ref{lad2})), let us first reinterpret the quantality using the zero point velocity $v_0$ (still in the $a$ $\sim$ r$_0$ approximation):

\begin{equation}
v_0=\frac{p_0}{m_N}\simeq\frac{\hbar}{m_Nr_0},
\end{equation}

and the quantality (\ref{la}) becomes:

\begin{equation}
\Lambda = \frac{\hbar^2}{m_Nr_0^2V'_0}\simeq\frac{\hbar v_0}{r_0V'_0}=\frac{\beta_0}{\alpha}
\label{laqf}
\end{equation}

with $\beta_0\equiv v_0$/c.

The quantality is therefore the ratio of the zero point motion velocity to the coupling constant. A large value of the quantality corresponds to a constituent
having a large zero point motion velocity compared to the effect of the attractive interaction and therefore delocalisation effects are important (if T$_N \sim$ V'$_0$), as
discussed in section \ref{sec:gll}. 

Let us now derive the expression of $\Lambda_{-2,1}$ in a similar way:

\begin{equation}
\Lambda_{-2,1}=\frac{1}{\alpha\eta^2}=\frac{\hbar cV'_0}{r_0m_N^2c^4}=\frac{\beta_0V'_0}{m_Nc^2}=\frac{\beta_0}{\eta}
\label{lad2bis}
\end{equation}

$\Lambda_{-2,1}$ is therefore the ratio of the zero point motion velocity to the inertia parameter. A large value
of $\Lambda_{-2,1}$ correspond to a large zero point motion velocity and a small inertia parameter. This could be interpreted as
a quantum mobility quantity, namely the mobility of the system due to zero point motion. 

The typical values of the quantum mobility for various systems is given in Table I. In nuclei, important effects arising from quantum fluidity are expected.

\section{Conclusion}

Several dimensionless ratios have been considered including the quantality and the localisation parameter in order to study various properties in finite many-body systems. Relevant lengthscales related to the localisation of the constituents have been identified. The role of delocalisation on the saturation density has been clarified using both the quantality and the localisation parameter. The emergence of clusters in low-density nuclear systems has been also described. Using an analytical mean-field description, the impact of the hard-core of the interaction on the nucleon delocalisation is analysed. The explicit dependence of the spin-orbit effect on the quantality of the system and on the coupling constant of the interaction has been provided. A general study of localisation in many-body systems is also undertaken investigating both the magnitude of relativistic effects and the low kinetic energy case.  In its most general derivation, localisation is not only driven by the quantality but rather by the quantum fluidity. A systematic search for dimensionless ratios is also undertaken, allowing to
reach the concepts of quantum fluidity and mobility.

The present study sheds light on the respective role of the quantality, the localisation parameter, the
coupling constant and the spin-orbit related parameter in many-body systems,
including finite ones. These are
a powerful tool to predict and compare the general behavior of various
many-body systems such as nuclei, electrons in atoms, or crystals. Many other properties of such systems could be further studied in the present framework, such as the 
binding energy and the shell effects, or a more detailed study on the dispersion of the wavefunction of the constituents.

\appendix

\section{The reduced Compton wavelength}

For a particle of mass m$_N$ the reduced Compton wavelength is defined as:

\begin{equation}
r_N \equiv \frac{\hbar c}{m_Nc^2}
\label{compt2}
\end{equation}

In the case of nucleons, r$_N \simeq$ 0.2 fm which is of the order of the interaction hard-core c $\sim$ 0.5 fm.
In the case of electrons, $r_N$=4.10$^{-3} $\AA\hspace{0.3mm}.

In the $a$ $\sim$ $r_0$ approximation, Eq. (\ref{key}) yields the following expression for the coupling
constant:

\begin{equation}
\alpha=\frac{1}{\Lambda}\frac{r_N}{r_0}
\label{cc2}
\end{equation}

Eq. (\ref{cc2}) shows that the coupling constant can be interpreted as the ratio of the
reduced Compton wavelength to the range of the interaction, normalised
by the quantality. In the case of nuclei ($\alpha\sim\Lambda\sim\eta\sim$1), the range of the interaction becomes 
of the same order than the reduced Compton wavelength.

Other dimensionless ratios (\ref{eta}), (\ref{la}) can also be expressed using the Compton wavelength:

\begin{equation}
\eta=\frac{\hbar c}{r_NV'_0}
\label{etarn}
\end{equation}

\begin{equation}
\Lambda =\frac{\hbar cr_N}{r_0^2V'_0}
\label{larn}
\end{equation}

The usual wavelength $\lambda_N$ of a given particle follows the De Broglie law:

\begin{equation}
\lambda_N=\frac{h}{p_N}=\frac{2\pi\hbar c}{p_N c}
\label{dbr}
\end{equation}

The special relativity provides the total energy of the particle:

\begin{equation}
E_N^2=p_N^2 c^2+m_N^2c^4=(m_Nc^2+T_N)^2
\label{spr}
\end{equation}

where T$_N$ is the kinetic energy of the particle

Eq. (\ref{spr}) implies:

\begin{equation}
p_N^2 c^2=T_N^2+2T_Nm_Nc^2
\label{rel}
\end{equation}

Using Eq. (\ref{rel}) and (\ref{dbr}) leads to

\begin{equation}
\lambda_N=\frac{2\pi}{\sqrt{\gamma^2-1}}r_N
\label{co2}
\end{equation}

where $\gamma$ here is the usual Lorentz factor (unrelated to (\ref{eq:gam})).

Eq. (\ref{co2}) shows that $\lambda_N$=r$_N$ if $\gamma\simeq$6.4,
that is if T$_N \simeq$ T$_C\equiv$5m$_N$c$^2$. 

Therefore, 

\begin{itemize}

\item if T$<$ T$_C$ then $\lambda_N$ $>$ r$_N$. Due to the large value
of T$_C$ this is almost always the case, except for ultra-relativistic particles. 
Taking T $\sim$ V'$_0$ and using Eq. (\ref{co2}) gives $\lambda_N \sim $ 30r$_N$ for nuclei and
about 600r$_N$ for electrons in atoms for instance. It corresponds a few Fermis for nucleons and to the 
Bohr radius for electrons in atoms. 

\item if T$=$ T$_C$ then $\lambda_N$ = r$_N$

\item if T$>$ T$_C$ then $\lambda_N$ $<$ r$_N$. This corresponds to a
velocity of the particle v$_N$$>$0.98 c, which is rarely the case in
bound many-body systems. 

\end{itemize}

It appears that the reduced Compton wavelength r$_N$ corresponds to
an almost minimal value of the constituent wavelength in a many-body
system, which could be reached with large kinetic factors. As mentioned above,
in the case of nucleons, it is close to the size of the hard-core of the interaction 
and to $r_l$ (Eq. (\ref{rlg})).

\section{Graphene}

Graphene can be described by masseless effective charge carriers
using a 2D Dirac equation \cite{nov05}. It should be noted that the discussion on the generalised localisation is valid 
irrespective of the dimension of the system. The effective fine structure constant in graphene is
$\alpha\simeq$2.5 \cite{shy07}. It is therefore relevant to
investigate whether the present approach encompasses the case of graphene.

In the case of masseless constituent (m$_N$=0), only two quantities
describe the system: V'$_0$ and (r$_0 \sim $ $a$) which are the magnitude and the range
of the effective interaction, respectively. Therefore, only one dimensionless
parameter can be built from these two quantities. In the case of
graphene, and following the present lines, it shall be
the coupling constant $\alpha$ (Eq. (\ref{cc})). 

The quantality becomes $\Lambda\rightarrow\infty$ in the case of graphene
because m$_N$=0. This means that graphene behaves likes a
``super''-QL, and that the wave functions of the charge careers shall be
much delocalised. This is in agreement with theoretical studies on
systems obeying to 2D Dirac equations \cite{zie98}.

The inertia parameter becomes $\eta$=0 in the case of graphene.
This specific case means that the LS effect vanishes (Eq. (\ref{x2})) in such a system, even if the
system would be of finite size. It should be noted that the key relation
(\ref{key}) still holds here:

\begin{equation}
\eta\Lambda\alpha^2=1
\end{equation}

Hence the graphene case can be described by the present
approach, as a limit case where $\eta$=0, $\Lambda$=$\infty$ and
$\alpha$ remains finite.

In the ultra-relativistic limit ($\eta$ $\sim$ 0), the generalised localisation parameter Eq. (\ref{delred}) then reduces to, using 
Eq. (\ref{key})

\begin{equation}
\frac{\lambda_N}{r_0}=2\pi\alpha^{-1}
\label{delur}
\end{equation}

In this case, the interaction coupling constant drives the localisation
behavior of the system rather than the quantality, at variance with
Eq. (\ref{eq:locla}). This shall be the case for the graphene.

\section*{Acknowledgement}
The authors thank M. Chabot, R. Lasseri and A. Mutschler for useful discussions.
P. Schuck and D. Vretenar are also thanked for reading the manuscript and useful comments.

\end{document}